\documentclass[journal]{IEEEtran}
\usepackage[T1]{fontenc}
\ifCLASSINFOpdf
\usepackage[pdftex]{graphicx}
\DeclareGraphicsExtensions{-eps-converted-to.pdf,.pdf,.jpeg,.png}
\else
\usepackage[dvips]{graphicx}
\fi
\usepackage[compress,nospace]{cite}
\usepackage[cmex10]{amsmath}
\usepackage{epstopdf,amsthm,stfloats,siunitx,amssymb,wasysym,algorithm,algorithmic,array,url, color,subfigure}
\usepackage[colorlinks,linkcolor=blue,anchorcolor=blue,citecolor=blue,bookmarks=false]{hyperref} 
\usepackage{footnote}
\makesavenoteenv{tabular}
\newtheorem{theorem}{Proposition}
\newtheorem{lemma}[theorem]{Lemma}

\begin{document}

\title{Off-Grid DOA Estimation Using Sparse Bayesian Learning in MIMO Radar With Unknown Mutual Coupling}

\author{Peng~Chen,~\IEEEmembership{Member,~IEEE}, Zhenxin~Cao,~\IEEEmembership{Member,~IEEE}, Zhimin~Chen,~\IEEEmembership{Member,~IEEE}, Xianbin~Wang,~\IEEEmembership{Fellow,~IEEE}
\thanks{This work was supported in part by the  National Natural Science Foundation of China (Grant No. 61801112, 61471117, 61601281), the Natural Science Foundation of Jiangsu Province (Grant No. BK20180357), the Open Program of State Key Laboratory of Millimeter Waves at Southeast University (Grant No. Z201804). \textit{(Corresponding author: Peng Chen)}}
\thanks{P.~Chen and Z.~Cao are with the State Key Laboratory of Millimeter Waves, Southeast University, Nanjing 210096, China (email: \{chenpengseu, caozx\}@seu.edu.cn).}
\thanks{Z.~Chen is with the School of Electronic and Information, Shanghai Dianji University, Shanghai 201306, China (email: chenzm@sdju.edu.cn).}
\thanks{X.~Wang is with the Department of Electrical and Computer Engineering, Western University, Canada (e-mail: xianbin.wang@uwo.ca).}
}

\markboth{Accepted by IEEE Transactions on Signal Processing}%
{Shell \MakeLowercase{\textit{et al.}}: Bare Demo of IEEEtran.cls for Journals}

\maketitle

\begin{abstract}
In the practical radar with multiple antennas, the antenna imperfections degrade the system performance. In this paper, the problem of estimating the direction of arrival (DOA) in multiple-input and multiple-output (MIMO) radar system with unknown mutual coupling effect between antennas is investigated. To exploit the target sparsity in the spatial domain, the compressed sensing (CS)-based methods have been proposed by discretizing the detection area and formulating the dictionary matrix, so an \emph{off-grid} gap is caused by the discretization processes. In this paper, different from the present DOA estimation methods, both the off-grid gap due to the sparse sampling and the unknown mutual coupling effect between antennas are considered at the same time, and a novel sparse system model for DOA estimation is formulated. Then, a novel sparse Bayesian learning (SBL)-based method named sparse Bayesian learning with the mutual coupling (SBLMC) is proposed, where an expectation-maximum (EM)-based method is established to estimate all the unknown parameters including the noise variance, the mutual coupling vectors, the off-grid vector and the variance vector of scattering coefficients. Additionally, the prior distributions for all the unknown parameters are theoretically derived. With regard to the DOA estimation performance, the proposed SBLMC method can outperform state-of-the-art methods in the MIMO radar with unknown mutual coupling effect, while keeping the acceptable computational complexity. 
\end{abstract}

\begin{IEEEkeywords}
	Compressed sensing, DOA estimation, MIMO radar, sparse Bayesian learning, mutual coupling.
\end{IEEEkeywords}

\section{Introduction} \label{sec1}
\IEEEPARstart{U}{nlike} the traditional phased-array radar, multiple-input and multiple-output (MIMO) radar systems can transmit correlated or uncorrelated signals and improve the degree of freedom, so the recent advancement of radar technology has directly led to MIMO radar systems. Usually, the MIMO radar systems can be classified into the colocated radar and distributed radar. In the colocated MIMO radar, the space between antennas is comparable with the wavelength of transmitted signals~\cite{Jian2007,pengTSP,davis2014}, such that the waveform diversity can be used to improve the target estimation performance. In the distributed MIMO radar, the distances between antennas are significant, so the spatial diversity of target's radar cross section (RCS) provided by the different view-angles of antennas can be used to improve the target detection performance~\cite{Haimovich2008,Chen:2017ena}. In general, the operation of distributed MIMO radar could be challenging due to the coordination and signal exchange among different antennas. Therefore, in this paper, a colocated MIMO radar system is investigated to estimate the directions of arrival (DOAs) for multiple targets. 

Traditionally, the DOA estimation can be achieved based on the discrete Fourier transform (DFT) of the received signal in the spatial domain~\cite{Veen1988}, but the resolution of such technique is too low to estimate multiple targets using one beam. The maximum likelihood-based and the subspace-based methods have been proposed to improve the DOA estimation performance, including multiple signal classification (MUSIC) method~\cite{ralph1986,schmidt1981}, Root-MUSIC method~\cite{Zoltowski1993}, and estimating signal parameters via rotational invariance techniques (ESPRIT) method~\cite{roy1989}. Additionally, the beamspace-based methods have also been proposed for DOA estimation~\cite{DU2009567}. For example, a beamspace design method is proposed in~\cite{6709762} for the DOA estimation in the MIMO radar with colocated antennas; a two-dimensional joint transmit array interpolation and beamspace design for planar array mono-static MIMO radar is proposed in~\cite{7962202} for  DOA estimation via tensor modeling; a transmit beamspace energy focusing method is proposed in~\cite{5728938} for MIMO radar with application to direction finding. Moreover, a combined Capon and approximate maximum likelihood (CAML) method is proposed in~\cite{4655353} for the estimation of target locations and amplitudes in MIMO radar. The tensor algebra and multidimensional harmonic retrieval are investigated for the DOA estimation of MIMO radar~\cite{5510180}, and an iterative adaptive Kronecker beamformer for MIMO radar is proposed in~\cite{5432999}. However, in the subspace-based DOA estimation methods, only the power of received signals from targets are exploited to establish the target and noise subspaces.

To exploit the target sparsity in the spatial domain,  compressed sensing (CS)-based methods are utilized to estimate DOA~\cite{Yang20166,yang2017SP,yao2011,Matteo2013,Matteo2016,Qing2016,6657792,6867380}. For example, in~\cite{DU2009567}, an iterative adaptive approach (IAA), maximum likelihood-based IAA (IAA-ML) and multi-snapshot sparse Bayesian learning (M-SBL) are given for the beamforming design based on the sparsity. The Bayesian approach together with expectation-maximization (EM) is used in M-SBL to realize the user parameter-free method. In~\cite{Tipping2001}, both the SBL and the relevance vector machine (RVM) are proposed, and the sparse reconstruction theory based on SBL is developed. In~\cite{shihao2008}, Bayesian compressive sensing (BCS) is developed for the sparse signal reconstruction with the CS measurements. In the CS-based method, the DOA estimation performance can be improved by the dense sampling grids. However, both the computational complexity and the mutual coherence between the columns in the dictionary are increased by the dense sampling grids. To improve the DOA estimation performance without the dense sampling grids, the \emph{off-grid} DOA estimation method is proposed in~\cite{Hao2011}. To further improve the sparse estimation performance, an off-grid sparse Bayesian inference (OGSBI) method is first proposed in~\cite{yang2013} for the DOA estimation. Then, by solving a specific polynomial in the off-grid DOA estimation problem, a Root-SBL method with low computational complexity is proposed in~\cite{jisheng2017}. In~\cite{Xiaohuan2016}, the perturbed SBL-based algorithm is proposed for the DOA estimation. A dictionary learning algorithm for off-grid sparse reconstruction is proposed in~\cite{Hojatollah2016}. In~\cite{qianli2018}, a grid evolution method is proposed to refine the grids for the SBL-based DOA estimation.

However, in the practical MIMO radar system, the mutual coupling effect between antennas cannot be ignored~\cite{Zhidong2012,Clerckx2007}. Therefore, the DOA estimation methods with the unknown mutual coupling effect have been proposed~\cite{Jianyan2017,Paolo2017,matthew2017}. Usually, the mutual coupling effects among the antennas can be characterized by a mutual coupling matrix, which is a symmetric Toeplitz matrix~\cite{liao2012,Thomas2018,ce2017}.
However, in the present papers, the effects of both off-grid in the CS-based method and the mutual coupling among antennas have not been considered simultaneously, especially, for the methods based on the Bayesian theory. 

In this paper, the DOA estimation problem in the MIMO radar system with unknown mutual coupling effect is investigated. Different from the present methods, a novel sparse-based system model considering both the off-grid gap and the unknown mutual coupling effect is formulated. Then, a novel estimation method named SBL with the mutual coupling (SBLMC) is proposed, where an EM-based method is established to iteratively estimate all the unknown parameters including the noise variance, the mutual coupling vectors, the off-grid vector and the variance vector of scattering coefficients. Additionally, we theoretically derive the prior distributions for all the unknown parameters including the target scattering coefficients, the mutual coupling vectors, the off-grid vector and the noise variance. Then, the proposed SBLMC is compared with state-of-the-art methods. To summarize, we make the contributions as follows:
\begin{itemize}
	\item \textbf{The Sparse-based model for MIMO radar with unknown mutual coupling effect:} Considering both the off-grid effect and mutual coupling effect, a novel system model of MIMO radar is formulated by exploiting the target sparsity in the spatial domain, so the DOA estimation problem is converted into a sparse reconstruction problem. 
            \item \textbf{The SBL-based method for DOA estimation with unknown mutual coupling effect:} A novel SBL-based method (SBLMC) is proposed for the DOA estimation in the MIMO radar system with unknown mutual coupling effect and off-grid effect. By estimating all the unknown parameters iteratively, the better estimation performance can be achieved than state-of-the-art methods.
            \item \textbf{The theoretical expressions for all unknown parameters in the SBL-based method:} In the proposed SBL-based method (SBLMC), the estimation expressions for all unknown parameters including the noise variance, the mutual coupling vectors, the variance vector of scattering coefficients and the off-grid vector are all theoretically derived.  
\end{itemize}

The remainder of this paper is organized as follows. The MIMO radar model for DOA estimation is elaborated in Section~\ref{sec2}. The proposed DOA estimation method with unknown mutual coupling, i.e., SBLMC, is presented in Section~\ref{sec3}.  Section~\ref{sec4} gives the simulation results. Finally, Section~\ref{sec5} concludes the paper.

\textit{Notations:} 
Matrices are denoted by capital letters in boldface (e.g., $\boldsymbol{A}$), and vectors are denoted by lowercase letters in boldface (e.g., $\boldsymbol{a}$). $\boldsymbol{I}_N$ denotes an $N\times N$ identity matrix. 
$\mathcal{E}\left\{\cdot\right\}$ denotes the expectation operation.
$\mathcal{CN}\left(\boldsymbol{a},\boldsymbol{B}\right)$ denotes the complex Gaussian distribution with the mean being $\boldsymbol{a}$ and the variance matrix being $\boldsymbol{B}$.  $\|\cdot\|_F$, $\|\cdot\|_2$, $\otimes$,  $\operatorname{Tr}\left\{\cdot\right\}$, $\operatorname{vec}\left\{\cdot\right\}$,  $(\cdot)^*$, $(\cdot)^\text{T}$ and $(\cdot)^\text{H}$ denote the Frobenius norm, the $\ell_2$ norm, the Kronecker product,  the trace of a matrix, the vectorization of a matrix, the conjugate, the matrix transpose and the  Hermitian transpose, respectively. $\mathbb{C}^{M\times N}$ denotes the set of $M\times N$ matrices with the entries being complex numbers. $\mathcal{R}\{a\}$ denotes the real part of complex value $a$. For a vector $\boldsymbol{a}$, $\left[\boldsymbol{a}\right]_n$ denotes the $n$-th entry of $\boldsymbol{a}$, and $\operatorname{diag}\{\boldsymbol{a}\}$ denotes a diagonal matrix with the diagonal entries from $\boldsymbol{a}$. For a matrix $\boldsymbol{A}$, $\left[\boldsymbol{A}\right]_n$ denotes the $n$-th column of $\boldsymbol{A}$, and $\operatorname{diag}\{\boldsymbol{A}\}$ denotes a vector with the entries from the diagonal entries of $\boldsymbol{A}$.

\section{MIMO Radar Model for DOA Estimation}\label{sec2}

\begin{figure}
	\centering
	\includegraphics[width=2.3in]{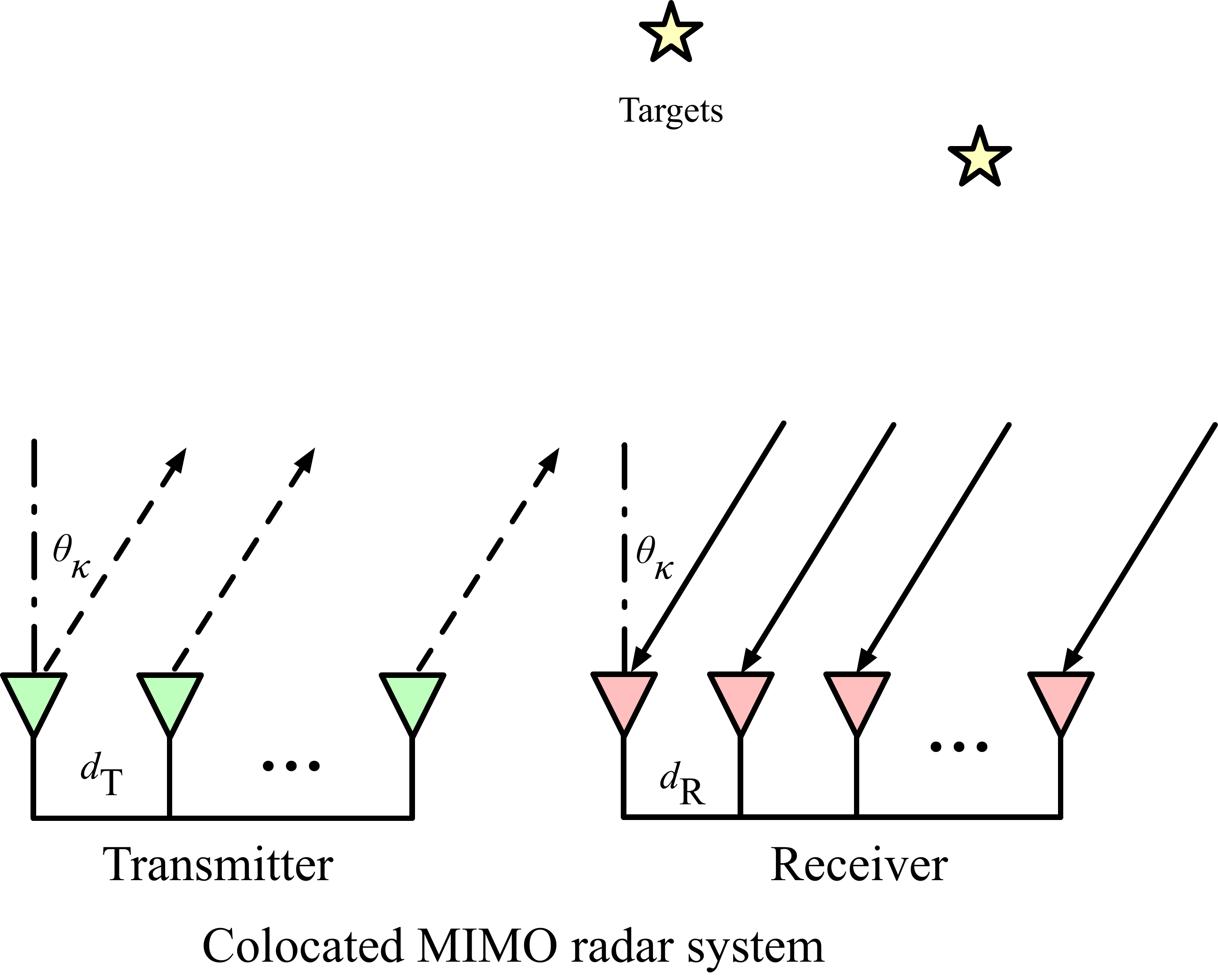}
	\caption{The MIMO radar system for DOA estimation.}
	\label{system}
\end{figure} 

As shown in Fig.~\ref{system}, we consider a colocated MIMO radar system, where $M$ transmitting antennas and $N$ receiving antennas are adopted. In the MIMO radar system, the orthogonal signals are transmitted by the antennas, and the waveform in the $m$-th ($m = 0,1,\dots, M-1$) transmitting antenna is $s_m(t)$. Assuming that $K$ far-field point targets in the same range cell are detected, we will consider the DOA estimation problem for these targets. The angle of the $k$-th ($k=0,1,\dots,K-1$) target is denoted as $\theta_k$. Therefore, under the assumption of narrowband signals, the received signals during the $p$-th ($p=0,1,\dots,P-1$, and $P$ denotes the number of pulses) pulse can be expressed as
\begin{align}
\boldsymbol{y}_{p}(t) &= 
\sum_{k=0}^{K-1}\gamma_{k,p}\boldsymbol{C}_\text{R}\boldsymbol{b}(\theta_k)
\left[\boldsymbol{C}_\text{T}\boldsymbol{a}(\theta_k)\right]^\text{T} \boldsymbol{s}(t-\tau_{\text{T}}-\tau_{\text{R}})
+\boldsymbol{v}_p(t)\notag\\
&\qquad (\tau_{\text{T}}+\tau_{\text{R}}\leq t \leq \tau_{\text{T}}+\tau_{\text{R}}+ T_{\text{P}}),
\end{align} 
where $T_{\text{P}}$ denotes the pulse duration, $\boldsymbol{v}_p(t)\triangleq \begin{bmatrix}v_{p,0}(t),v_{p,1}(t),\dots,v_{p,N-1}(t)\end{bmatrix}^{\text{T}}$ denotes the additive white Gaussian noise (AWGN), and $\gamma_{k,p}$ denotes the scattering coefficient of the $k$-th target during the $p$-th pulse. $\tau_{\text{T}}$ denotes the signal propagation time delay between the transmitter and the range cell, and $\tau_{\text{R}}$ denotes the delay between the range cell and the receiver. The received signals and transmitted signals are respectively defined as
\begin{align}
\boldsymbol{y}_p(t)&\triangleq \begin{bmatrix}y_0(t),y_1(t),\dots,y_{N-1}(t)\end{bmatrix}^{\text{T}},\\
\boldsymbol{s}(t)&\triangleq \begin{bmatrix}
s_0(t),s_1(t),\dots,s_{M-1}(t)
\end{bmatrix}^{\text{T}}.
\end{align}
The steering vectors of the transmitter and receiver are respectively denoted as
\begin{align}
\boldsymbol{a}(\theta)&\triangleq \begin{bmatrix}
1, e^{j2\pi\frac{d_\text{T}}{\lambda}\sin\theta},\dots,e^{j2\pi\frac{(M-1)d_\text{T}}{\lambda}\sin\theta}
\end{bmatrix}^\text{T},\\
\boldsymbol{b}(\theta)&\triangleq \begin{bmatrix}
1, e^{j2\pi\frac{d_\text{R}}{\lambda}\sin\theta},\dots,e^{j2\pi\frac{(N-1)d_\text{R}}{\lambda}\sin\theta}
\end{bmatrix}^\text{T},
\end{align}
where $d_\text{T}$ and $d_\text{R}$ denote the distance separation between two neighboring antennas  in the transmitter and receiver, respectively, and $\lambda$ denotes the wavelength.
$\boldsymbol{C}_\text{T}\in\mathbb{C}^{M\times M}$ and $\boldsymbol{C}_\text{R}\in\mathbb{C}^{N\times N}$ denotes the mutual coupling matrices in the transmitter and receiver, respectively. The mutual coupling matrix $\boldsymbol{C}_\text{T}$ is a symmetric Toeplitz matrix, and can be expressed as~\cite{ce2017} 
\begin{align}
	\boldsymbol{C}_\text{T}=
	\begin{bmatrix}
	1 & c_{\text{T},1} &\dots & c_{\text{T},M-1}\\
	c_{\text{T},1} & 1 & \dots & c_{\text{T},M-2}\\
	\vdots & \vdots & \ddots & \vdots\\
	c_{\text{T},M-1} & \dots & c_{\text{T},1} & 1
	\end{bmatrix},
\end{align}
where $c_{\text{T},m}$ denotes the $m$-th entry of a vector $\boldsymbol{c}_{\text{T}}\triangleq \begin{bmatrix}
c_{\text{T},0}, c_{\text{T},1},\dots, c_{\text{T},M-1}
\end{bmatrix}^{\text{T}}$.
Alternatively, the entry of $\boldsymbol{C}_\text{T}$ at the $m$-th row and $m'$-th column can be also written as
\begin{align}
C_{\text{T},m,m'}=\begin{cases}
1,&m=m'\\
c_{\text{T},|m-m'|},&\text{otherwise}
\end{cases}.
\end{align}
Using the same method, we can obtain the expression of $\boldsymbol{C}_\text{R}$.

Since the orthogonal signals are adopted in the transmitting antennas, we can use $M$ matched filters corresponding to the $M$ orthogonal signals to distinguish the orthogonal signals. We estimate the parameters of targets at a specific range cell, so the delays $\tau_{\text{T}}$ and $\tau_{\text{R}}$ are omitted. Therefore, passing the $m$-th matched filter (designed for the $m$-th signal)~\cite{Jian2007}, the received signals $\boldsymbol{r}_{p}(t)$ from the same range cell are sampled at $T_{\text{P}}$ and obtained as
\begin{align}
\boldsymbol{r}_{p,m}&\triangleq \sum_{k=0}^{K-1}\gamma_{k,p}\boldsymbol{C}_\text{R}\boldsymbol{b}(\theta_k)
\left[\boldsymbol{C}_\text{T}\boldsymbol{a}(\theta_k)\right]^\text{T} 
\underbrace{\begin{bmatrix}
\int s_0(t)s_m^{\text{H}}(t)dt\\
\vdots\\
\int s_m(t)s_m^{\text{H}}(t)dt\\
\vdots\\
\int s_{M-1}(t)s_m^{\text{H}}(t)dt
\end{bmatrix}}_{\boldsymbol{e}^M_m}\notag\\
&\qquad
+\underbrace{\begin{bmatrix}
	\int v_{p,0}(t)s_m^{\text{H}}(t)dt\\
	\vdots\\
	\int v_{p,m}(t)s_m^{\text{H}}(t)dt\\
	\vdots\\
	\int v_{p,M-1}(t)s_m^{\text{H}}(t)dt
	\end{bmatrix}}_{\boldsymbol{n}_{p,m}}\\
&=\sum_{k=0}^{K-1}\gamma_{k,p}\boldsymbol{C}_\text{R}\boldsymbol{b}(\theta_k)
\left[\boldsymbol{C}_\text{T}\boldsymbol{a}(\theta_k)\right]^\text{T} \boldsymbol{e}^M_m+\boldsymbol{n}_{p,m}\notag\\
&=
\sum_{k=0}^{K-1}\gamma_{k,p}\left[\boldsymbol{C}_\text{T}\boldsymbol{a}(\theta_k)\right]_m\boldsymbol{C}_\text{R}\boldsymbol{b}(\theta_k)+\boldsymbol{n}_{p,m},\notag
\end{align}
where $\boldsymbol{e}^M_m$ is a $M\times 1$ vector with  the $m$-th entry being $1$ and others entries being zeros, and $\boldsymbol{n}_{p,m}$ is the additive noise. Collect $\boldsymbol{r}_{p,m}$ into a matrix, and we can obtain
\begin{align}
\boldsymbol{R}_p&\triangleq\begin{bmatrix}
\boldsymbol{r}^\text{T}_{p,0}\\
\boldsymbol{r}^\text{T}_{p,1}\\
\vdots\\
\boldsymbol{r}^\text{T}_{p,M-1}
\end{bmatrix}  = \sum_{k=0}^{K-1}\gamma_{k,p}
\boldsymbol{C}_\text{T}\boldsymbol{a}(\theta_k)
\left[\boldsymbol{C}_\text{R}\boldsymbol{b}(\theta_k)\right]^\text{T}+\boldsymbol{N}_p
\end{align}
where the noise matrix is defined as $\boldsymbol{N}_p\triangleq \begin{bmatrix}
\boldsymbol{n}_{p,0},\boldsymbol{n}_{p,1},\dots,\boldsymbol{n}_{p,M-1}
\end{bmatrix}^\text{T}$. Vectorizing the receiving signal matrix into a vector $\boldsymbol{r}_p\triangleq\operatorname{vec}\left\{\boldsymbol{R}_p\right\}$, we can obtain
\begin{align}
\boldsymbol{r}_p&=\sum_{k=0}^{K-1}\gamma_{k,p}\operatorname{vec}\left\{
\boldsymbol{C}_\text{T}\boldsymbol{a}(\theta_k)
\left[\boldsymbol{C}_\text{R}\boldsymbol{b}(\theta_k)\right]^\text{T}\right\}+\boldsymbol{n}_p\\
& = \sum_{k=0}^{K-1}\gamma_{k,p}\left[\boldsymbol{C}_\text{R}\boldsymbol{b}(\theta_k)\right]\otimes\left[\boldsymbol{C}_\text{T}\boldsymbol{a}(\theta_k)\right]+\boldsymbol{n}_p,\notag
\end{align}
where  $\boldsymbol{n}_p\triangleq \operatorname{vec}\left\{\boldsymbol{N}_p\right\}$.

Alternatively, the received signal $\boldsymbol{r}_p$ can be also rewritten into a matrix form 
\begin{align}
\boldsymbol{r}_p=\boldsymbol{\Delta\gamma}_p+\boldsymbol{n}_p,
\end{align}
where $\boldsymbol{\gamma}_p\triangleq\begin{bmatrix}
\gamma_{p,0},\gamma_{p,1},\dots,\gamma_{p,K-1}
\end{bmatrix}^\text{T}$, $\boldsymbol{\Delta}\triangleq \begin{bmatrix}\boldsymbol{\delta}_0,\boldsymbol{\delta}_1,\dots,\boldsymbol{\delta}_{K-1}\end{bmatrix}$, and 
\begin{align}
\boldsymbol{\delta}_k & \triangleq \left[\boldsymbol{C}_\text{R}\boldsymbol{b}(\theta_k)\right]\otimes\left[\boldsymbol{C}_\text{T}\boldsymbol{a}(\theta_k)\right]\\
& = \left[\boldsymbol{C}_\text{R}\otimes \boldsymbol{C}_\text{T}\right]\left[\boldsymbol{b}(\theta_k)\otimes\boldsymbol{a}(\theta_k)\right].\notag
\end{align}
By defining $\boldsymbol{C}\triangleq \boldsymbol{C}_\text{R}\otimes \boldsymbol{C}_\text{T}$ and $\boldsymbol{d}(\theta_k)=\boldsymbol{b}(\theta_k)\otimes\boldsymbol{a}(\theta_k)$, we have
\begin{align}\boldsymbol{\Delta}&= \boldsymbol{C}\begin{bmatrix}\boldsymbol{d}(\theta_0),\boldsymbol{d}(\theta_1),\dots,\boldsymbol{d}(\theta_{K-1})\end{bmatrix} =\boldsymbol{C}\boldsymbol{D},
\end{align}
where $\boldsymbol{D}\triangleq \begin{bmatrix}\boldsymbol{d}(\theta_0),\boldsymbol{d}(\theta_1),\dots,\boldsymbol{d}(\theta_{K-1})\end{bmatrix}$. Therefore, the received signal with mutual coupling effect can be formulated by the following model
\begin{align}\label{EQrecv}
\boldsymbol{r}_p & =\boldsymbol{CD}\boldsymbol{\gamma}_p+\boldsymbol{n}_p.
\end{align}

To simplify the formula with the mutual coupling matrix in (\ref{EQrecv}), we will use the following lemma:
\begin{lemma}\label{lemma1}
For complex symmetric Toeplitz matrix $\boldsymbol{A}=\operatorname{Toeplitz}\left\{\boldsymbol{a}\right\}\in\mathbb{C}^{M\times M}$ and complex vector $\boldsymbol{c}\in\mathbb{C}^{M\times 1}$, we have~\cite{Termos:2004ch,LIU2012517,ce2017}
\begin{equation}
\boldsymbol{Ac}=\boldsymbol{Qa},
\end{equation}	
where $\boldsymbol{a}$ is a vector formed by the first row of $\boldsymbol{A}$, and $\boldsymbol{Q}=\boldsymbol{Q}_1+\boldsymbol{Q_2}$ with the $p$-th ($p=0,1,\dots, M-1$) row and $q$-th ($q=0,1,\dots,M-1$) column entries being
\begin{align}
[\boldsymbol{Q}_1]_{p,q}&=\begin{cases}
c_{p+q},&p+q\leq M-1\\
0,&\text{otherwise}
\end{cases},\\
[\boldsymbol{Q}_2]_{p,q}&=\begin{cases}
	c_{p-q},&p\geq q\geq 1\\
	0,&\text{otherwise}
\end{cases}.
\end{align}
\end{lemma}

Based on Lemma~\ref{lemma1}, $\boldsymbol{\delta}_k$ can be rewritten as
\begin{align}
\boldsymbol{\delta}_k &= \left[\boldsymbol{C}_\text{R}\boldsymbol{b}(\theta_k)\right]\otimes\left[\boldsymbol{C}_\text{T}\boldsymbol{a}(\theta_k)\right]\\
& =  \left[
\boldsymbol{Q}_b(\theta_k)\boldsymbol{c}_\text{R}\right]\otimes\left[
\boldsymbol{Q}_a(\theta_k)\boldsymbol{c}_\text{T}\right]\notag\\
& =  \left[
\boldsymbol{Q}_b(\theta_k)\otimes \boldsymbol{Q}_a(\theta_k)\right]\boldsymbol{c},\notag 
\end{align}
where $\boldsymbol{c}\triangleq\boldsymbol{c}_\text{R}\otimes \boldsymbol{c}_\text{T}$, and the $m$-th entry of $\boldsymbol{c}_\text{T}$ and the $n$-th entry of $\boldsymbol{c}_\text{R}$ respectively are 
\begin{align}
[\boldsymbol{c}_{\text{T}}]_m&=\begin{cases}
1, &m=0\\
c_{\text{T},m}, & \text{otherwise}
\end{cases},\\
[\boldsymbol{c}_{\text{R}}]_n&=\begin{cases}
1, &n=0\\
c_{\text{R},n}, &\text{otherwise}
\end{cases}.
\end{align}
$\boldsymbol{Q}_a(\theta_k)$ and $\boldsymbol{Q}_b(\theta_k)$ can be obtained as 
\begin{align}
\boldsymbol{Q}_a(\theta_k)&=\boldsymbol{Q}_{a1}(\theta_k)+\boldsymbol{Q}_{a2}(\theta_k)\label{Qa},\\
\boldsymbol{Q}_b(\theta_k)&=\boldsymbol{Q}_{b1}(\theta_k)+\boldsymbol{Q}_{b2}(\theta_k)\label{Qb},
\end{align}
where the $p$-th row and $q$-th column entries of $\boldsymbol{Q}_{a1}(\theta_k)$, $\boldsymbol{Q}_{a2}(\theta_k)$, $\boldsymbol{Q}_{b1}(\theta_k)$ and $\boldsymbol{Q}_{b2}(\theta_k)$  respectively are
\begin{align}
[\boldsymbol{Q}_{a1}]_{p,q}&=\begin{cases}
[\boldsymbol{a}(\theta_k)]_{p+q},&p+q\leq M-1\\
0,&\text{otherwise}
\end{cases},\\
[\boldsymbol{Q}_{a2}]_{p,q}&=\begin{cases}
[\boldsymbol{a}(\theta_k)]_{p-q},&p\geq q\geq 1\\
0,&\text{otherwise}
\end{cases},\\
[\boldsymbol{Q}_{b1}]_{p,q}&=\begin{cases}
[\boldsymbol{b}(\theta_k)]_{p+q},&p+q\leq N-1\\
0,&\text{otherwise}
\end{cases},\\
[\boldsymbol{Q}_{b2}]_{p,q}&=\begin{cases}
[\boldsymbol{b}(\theta_k)]_{p-q},&p\geq q\geq 1\\
0,&\text{otherwise}
\end{cases}.
\end{align}

Therefore, we have
\begin{align}
\boldsymbol{\Delta} = \boldsymbol{Q}\left[\boldsymbol{I}_K\otimes \boldsymbol{c}\right],
\end{align}
where 
\begin{align}
\boldsymbol{Q}\triangleq \begin{bmatrix}
\boldsymbol{Q}_b(\theta_0)\otimes \boldsymbol{Q}_a(\theta_0),\dots,
\boldsymbol{Q}_b(\theta_{K-1})\otimes \boldsymbol{Q}_a(\theta_{K-1})
\end{bmatrix}.\label{Q}
\end{align}
Then, the received signal in (\ref{EQrecv}) can be rewritten as
\begin{align}\label{EQrecv2}
\boldsymbol{r}_p&=\boldsymbol{Q}\left(\boldsymbol{I}_K\otimes \boldsymbol{c}\right)\boldsymbol{\gamma}_p+\boldsymbol{n}_p\notag\\
&=\boldsymbol{Q}\left(\boldsymbol{\gamma}_p\otimes \boldsymbol{c}\right)+\boldsymbol{n}_p.
\end{align}

Collect the $P$ pulses into a matrix, and the received signal can be finally obtained as
\begin{align}
\boldsymbol{R}
&=\boldsymbol{Q}\begin{bmatrix}\boldsymbol{\gamma}_0\otimes \boldsymbol{c},\boldsymbol{\gamma}_1\otimes \boldsymbol{c},\dots,\boldsymbol{\gamma}_{P-1}\otimes \boldsymbol{c}\end{bmatrix}+\boldsymbol{N}\notag \\
&=\boldsymbol{Q}(\boldsymbol{\Gamma}\otimes \boldsymbol{c}) +\boldsymbol{N},
\end{align}
where $\boldsymbol{R}\triangleq\begin{bmatrix}
	\boldsymbol{r}_0,\boldsymbol{r}_1,\dots,\boldsymbol{r}_{P-1}
\end{bmatrix}$, $\boldsymbol{N}\triangleq\begin{bmatrix}
	\boldsymbol{n}_0,\boldsymbol{n}_1,\dots,\boldsymbol{n}_{P-1}
\end{bmatrix}$, $\boldsymbol{\Gamma}\triangleq\begin{bmatrix}
	\boldsymbol{\gamma}_0,\boldsymbol{\gamma}_1,\dots,\boldsymbol{\gamma}_{P-1}
\end{bmatrix}$, and the $u$-th row and $p$-th column of $\boldsymbol{\Gamma}$ is denoted as $\Gamma_{u,p}$. In this paper, we will estimate the DOAs from $\boldsymbol{R}$ with the unknown mutual coupling vector $\boldsymbol{c}$, the target scattering coefficients $\boldsymbol{\Gamma}$ and the noise variance $\sigma^2_n$.

\section{DOA Estimation Method With Unknown Mutual Coupling}\label{sec3}

\subsection{The Off-Grid Sparse Model}
Discretize the angle of detection area into $U$ grids $\boldsymbol{\zeta}\triangleq \begin{bmatrix}
	\zeta_0,\zeta_1,\dots, \zeta_{U-1}
\end{bmatrix}$, and the $u$-th discretized angle are denoted as $\zeta_u$. Then, a dictionary matrix can be formulated as
\begin{align}
	\boldsymbol{\Psi}\triangleq\begin{bmatrix}
		\boldsymbol{\Phi}(\zeta_0),\boldsymbol{\Phi}(\zeta_1),\dots,\boldsymbol{\Phi}(\zeta_{U-1})
	\end{bmatrix}\in\mathbb{C}^{MN\times UMN},
\end{align}
where $\boldsymbol{\Phi}(\zeta_u)\triangleq \boldsymbol{Q}_b(\zeta_u)\otimes \boldsymbol{Q}_a(\zeta_u)$, and the space between discretized angles, also known as grid size, is $\delta\triangleq |\zeta_{u+1}-\zeta_u|$. For the dictionary matrix $\boldsymbol{\Psi}$, the restricted isometry property (RIP) with constant $\delta_{\varpi}$ is defined as~\cite{1542412}
\begin{align}
(1-\delta_{\varpi})\|\boldsymbol{x}\|^2_2\leq \|\boldsymbol{\Psi}\boldsymbol{x}\|^2_2\leq (1+\delta_{\varpi})\|\boldsymbol{x}\|^2_2,
\end{align}
for all $\varpi$-sparse vector $\boldsymbol{x}$ ($\varpi=KMN$). If $\boldsymbol{x}$ is $\varpi$-sparse and $\boldsymbol{\Psi}$ satisfies $\delta_{2\varpi }+\delta_{3 \varpi}<1$, then $\boldsymbol{x}$ is the unique $\ell_1$ minimizer~\cite{4016283}. However, the tight RIP constant of a given matrix $\boldsymbol{\Psi}$ is difficult to compute, so we calculate the minimum DOA separation for multiple targets. As described in~\cite{Fernandez1}, in our scenario (the system parameters are given in Section~\ref{sec4}), the minimum DOA separation can be obtained as $\ang{9.67}$. 

However, for the $k$-th target, the DOA is $\theta_k$ and is not at the discretized grids exactly, so the sub-matrix for the $k$-th target can be approximated by
\begin{align}
	\boldsymbol{\Phi}(\theta_k)&=\boldsymbol{\Phi}(\zeta_{u_k}+(\theta_k-\zeta_{u_k}))\notag\\
	&\approx  \left[\boldsymbol{Q}_{b}(\zeta_{u_k})+(\theta_k-\zeta_{u_k})\left.\frac{\partial \boldsymbol{Q}_{b}(\zeta)}{\partial \zeta}\right|_{\zeta=\zeta_{u_k}}
	\right]\notag\\
	&\qquad \otimes  \left[\boldsymbol{Q}_{a}(\zeta_{u_k})+(\theta_k-\zeta_{u_k})\left.\frac{\partial \boldsymbol{Q}_{a}(\zeta)}{\partial \zeta}\right|_{\zeta=\zeta_{u_k}}
	\right]\notag\\
	&\approx  \boldsymbol{\Phi}(\zeta_{u_k}) +(\theta_k-\zeta_{u_k})\boldsymbol{\Omega}(\zeta_{u_k}),
\end{align}
where $\zeta_{u_k}$ is the discretized grid angle nearest to the target DOA $\theta_k$, and we define the first order of derivative as 
\begin{align}
	\boldsymbol{\Omega}(\zeta_{u_k})&\triangleq \boldsymbol{Q}_{b}(\zeta_{u_k})\otimes\left.\frac{\partial \boldsymbol{Q}_{a}(\zeta)}{\partial \zeta}\right|_{\zeta=\zeta_{u_k}} \notag\\
	&\quad+  \left.\frac{\partial \boldsymbol{Q}_{b}(\zeta)}{\partial \zeta}\right|_{\zeta=\zeta_{u_k}} \otimes \boldsymbol{Q}_{a}(\zeta_{u_k}).
\end{align}

By formulating a sparse matrix $\boldsymbol{X}\in\mathbb{C}^{U\times P}$ with the columns $\boldsymbol{x}_p$ ($p=0,1,\dots,P-1$) having the same support set, i.e., $\boldsymbol{X}\triangleq \begin{bmatrix}
	\boldsymbol{x}_0,\boldsymbol{x}_1,\dots,\boldsymbol{x}_{P-1}
\end{bmatrix}$, the received signal can be approximated by a sparse-based model
\begin{align}
	\boldsymbol{R}\approx\left[\boldsymbol{\Psi}+\boldsymbol{\Xi}\left(\operatorname{diag}\left\{
	\boldsymbol{\nu} \right\}\otimes \boldsymbol{I}_{MN}\right)\right](\boldsymbol{X}\otimes \boldsymbol{c})+\boldsymbol{N},
\end{align}
where $\boldsymbol{\Xi}\triangleq\begin{bmatrix}
	\boldsymbol{\Omega}(\zeta_0),\boldsymbol{\Omega}(\zeta_1),\dots,\boldsymbol{\Omega}(\zeta_{U-1})
\end{bmatrix}$, and the $u$-th sub-matrix can be also written as $\boldsymbol{\Xi}_u\triangleq \boldsymbol{\Omega}(\zeta_u)$ to simplify the notation. The $u$-th row and $p$-th column of sparse matrix $\boldsymbol{X}\in\mathbb{C}^{U \times P}$ is 
\begin{align}
	X_{u,p}=\begin{cases}
		\Gamma_{u_k,p},&u=u_k\\
		0,&\text{otherwise}
	\end{cases},
\end{align}
and the $u$-th entry of the off-grid vector $\boldsymbol{\nu}\in\mathbb{R}^{U\times 1}$ is
\begin{align}
	\nu_u=\begin{cases}
		\theta_k-\zeta_{u_k}, & u=u_k\\
		0, & \text{otherwise}
	\end{cases}.
\end{align}

Finally, by absorbing the approximation into the additive noise, the off-grid sparse model in the MIMO radar with unknown mutual coupling effect can be described by a sparse model
\begin{align}
	\boldsymbol{R}=\boldsymbol{\Upsilon}(\boldsymbol{\nu})(\boldsymbol{X}\otimes \boldsymbol{c}_\text{R}\otimes \boldsymbol{c}_\text{T})+\boldsymbol{N},
\end{align} 
where $\boldsymbol{\Upsilon}(\boldsymbol{\nu})\triangleq \boldsymbol{\Psi}+\boldsymbol{\Xi}\left(\operatorname{diag}\left\{
	\boldsymbol{\nu} \right\}\otimes \boldsymbol{I}_{MN}\right) $. With the received signal $\boldsymbol{R}$, we can estimate the target DOAs $\theta_k$ ($k=0,1,\dots,K-1$) with the unknown parameters including the sparse matrix $\boldsymbol{X}$, the off-grid vector $\boldsymbol{\nu}$, and the mutual coupling vectors $\boldsymbol{c}_\text{T}$ and $\boldsymbol{c}_\text{R}$. The DOAs can be obtained from the support sets of $\boldsymbol{X}$, the target scattering coefficients are obtained from the nonzero entries of $\boldsymbol{X}$, and the mutual coupling matrices can be obtained from $\boldsymbol{c}_\text{T}$ and $\boldsymbol{c}_\text{R}$.
	
\subsection{Sparse Bayesian Learning-Based DOA Estimation Method}

\begin{figure}
	\centering
	\includegraphics[width=3.5in]{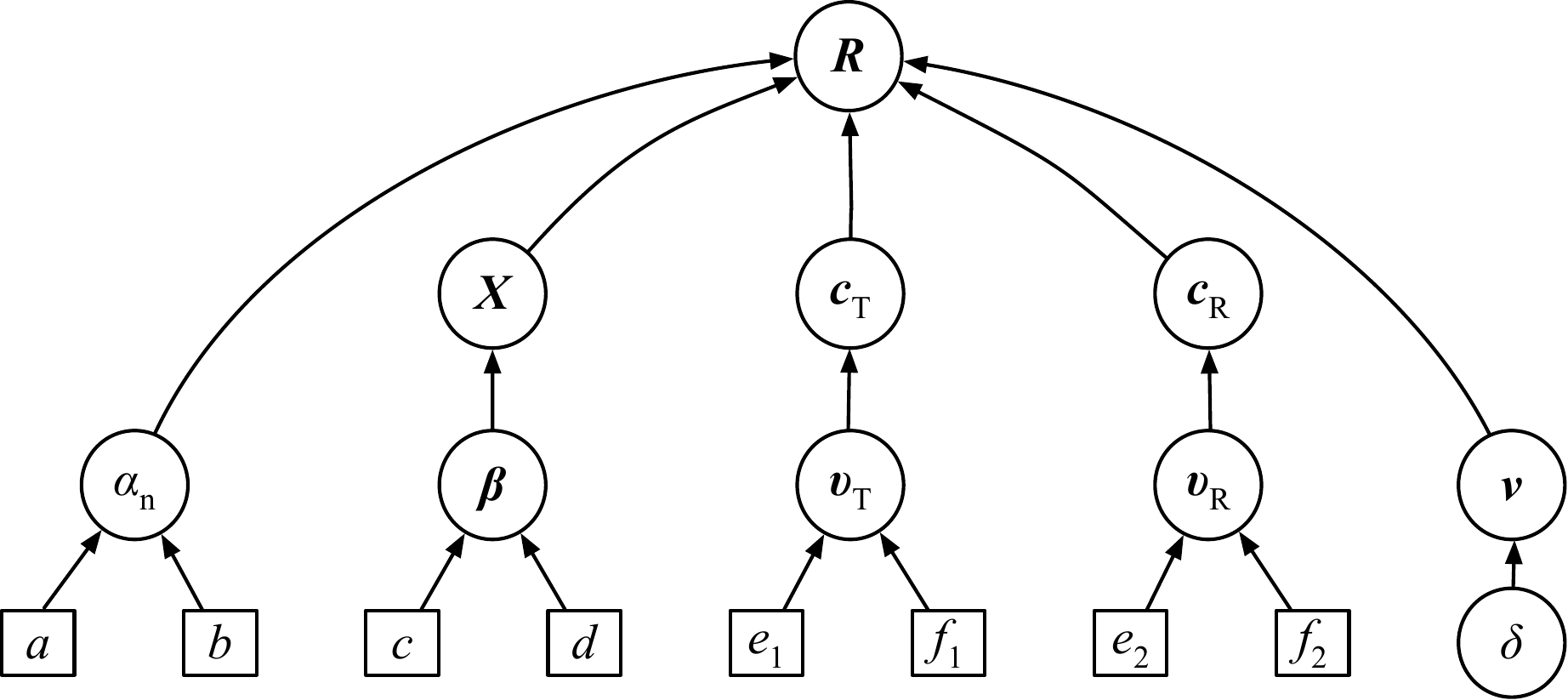}
	\caption{Graphical model of SBLMC (rectangles are the hyperparameters, circles are the radar parameters and signals). }
	\label{Bayesian}
\end{figure}

In this paper, we propose an SBL-based method to estimate the target DOAs with unknown mutual coupling effect, and the proposed method is named as SBL with the mutual coupling (SBLMC). The graphical model of SBLMC is given in Fig.~\ref{Bayesian}, where the unknown parameters are determined by the hyperparameters, and the received signal $\boldsymbol{S}$ is determined by radar parameters and signals. To realize the SBLMC algorithm, the distribution assumptions are given as follows.

We assume that the additive noise is white (circularly symmetric) Gaussian noise with the noise variance being $\sigma^2_n$, and the distribution of noise can be expressed as
\begin{align}
	p(\boldsymbol{N}|\sigma^2_n)=\prod^{U-1}_{u=0}\mathcal{CN}(\boldsymbol{n}_p|\boldsymbol{0}_{MN\times 1},\sigma^2_n\boldsymbol{I}_{MN}),
\end{align}
where the complex Gaussian distribution is defined as
\begin{align}
	\mathcal{CN}(\boldsymbol{x}|\boldsymbol{a},\boldsymbol{\Sigma})=\frac{1}{\pi^N\det(\boldsymbol{\Sigma})}e^{-(\boldsymbol{x}-\boldsymbol{a})^\text{H}\boldsymbol{\Sigma}^{-1}(\boldsymbol{x}-\boldsymbol{a})}.
\end{align}

When the noise variance $\sigma^2_n$ is unknown, by defining a hyperparamter, i.e., the \emph{precision}, $\alpha_n\triangleq \sigma^{-2}_n$, a Gamma distribution can be adopted to describe the inverse of noise variance
\begin{align}
	p(\alpha_n)=\mathfrak{G}(\alpha_n;a,b),
\end{align}
where $a$ and $b$ are the hyperparameters for $\alpha_n$, and
\begin{align}
	\mathfrak{G}(\alpha_n;a,b)&\triangleq \Gamma^{-1}(a)b^a\alpha_n^{a-1}e^{-b\alpha_n},\\
	\Gamma(a)&\triangleq \int^\infty_0x^{a-1}e^{-x}dx.
\end{align}
Note that the Gamma distribution $\alpha_n\sim \mathfrak{G}(\alpha_n;a,b)$ is a conjugate prior of the Gaussian distribution given mean with unknown variance $x\sim \mathcal{N}(x|0,\alpha^{-1}_n)$, so the posterior distribution $p(\alpha_n|x)$ also follows a Gamma distribution. Therefore,  the assumption of Gamma distribution for the \emph{precision} $\alpha_n$ can simplify the following analysis.

When the scattering coefficients $\boldsymbol{\Gamma}$ are independent among snapshots, we can also assume that the sparse matrix $\boldsymbol{X}$ follows a Gaussian distribution
\begin{align}
	p(\boldsymbol{X}|\boldsymbol{\Lambda}_x)=\prod_{p=0}^{P-1}\mathcal{CN}(\boldsymbol{x}_p|\boldsymbol{0}_{U\times 1}, \boldsymbol{\Lambda}_x),
\end{align}
where $\boldsymbol{\Lambda}_x\in\mathbb{R}^{U\times U}$ is a diagonal matrix with the $u$-th diagonal entry being $\sigma^2_{x,u}$. Usually, the sparseness prior is the Laplace density function~\cite{Tipping2001,shihao2008}, but the Laplace prior is not conjugate to the Gaussian likelihood. Therefore, for simplification, we use the Gaussian prior for the sparse matrix $\boldsymbol{X}$ and obtain the estimation expressions in closed form.
Then, by defining the \emph{precision} $\boldsymbol{\beta}\triangleq \begin{bmatrix}
	\beta_0,\beta_1,\dots,\beta_{U-1}
\end{bmatrix}^\text{T} $ and $\beta_u\triangleq \sigma^{-2}_{x,u}$, we have the following Gamma prior for $\boldsymbol{\beta}$ 
\begin{align}
	p(\boldsymbol{\beta};c,d)=\prod^{U-1}_{u=0}\mathfrak{G}(\beta_u;c,d),
\end{align} 
where $c$ and $d$ are the hyperparmaters for $\boldsymbol{\beta}$.

Similarly, when the mutual coupling coefficients are independent with antennas, we can also assume that the mutual coupling vectors $\boldsymbol{c}_\text{T}$ and $\boldsymbol{c}_\text{R}$ follow Gaussian distributions
\begin{align}
	p(\boldsymbol{c}_\text{T}|\boldsymbol{\Lambda}_\text{T})&=\prod^{M-1}_{m=0}\mathcal{CN}(c_{\text{T},m}|0,\sigma^2_{\text{T},m}),\\
	p(\boldsymbol{c}_\text{R}|\boldsymbol{\Lambda}_\text{R})&=\prod^{N-1}_{n=0}\mathcal{CN}(c_{\text{R},n}|0,\sigma^2_{\text{R},n}),
\end{align}
where $\boldsymbol{\Lambda}_\text{T}\in\mathbb{R}^{M\times M}$ is a diagonal matrix with the $m$-th diagonal entry being $\sigma^2_{\text{T},m}$, and $\boldsymbol{\Lambda}_\text{R}\in\mathbb{R}^{N\times N}$ is a diagonal matrix with the $n$-th diagonal entry being $\sigma^2_{\text{R},n}$. Define the \emph{precisions} $\boldsymbol{\vartheta}_\text{T}\triangleq\begin{bmatrix}
	\vartheta_{\text{T},0},\vartheta_{\text{T},1},\dots,\vartheta_{\text{T},M-1}
\end{bmatrix}^\text{T}$ ($\vartheta_{\text{T},m}\triangleq\sigma^{-2}_{\text{T},m}$) and $\boldsymbol{\vartheta}_\text{R}\triangleq \begin{bmatrix}
	\vartheta_{\text{R},0},\vartheta_{\text{R},1},\dots,\vartheta_{\text{R},N-1}
\end{bmatrix}^\text{T}$ ($\vartheta_{\text{R},n}\triangleq \sigma^{-2}_{\text{R},n}$). Then, we can have the following Gamma distributions
\begin{align}
	p(\boldsymbol{\vartheta}_\text{T};e_1,f_1)&=\prod^{M-1}_{m=0}\mathfrak{G}(\vartheta_{\text{T},m};e_1,f_1),\\
	p(\boldsymbol{\vartheta}_\text{R};e_2,f_2)&=\prod^{N-1}_{n=0}\mathfrak{G}(\vartheta_{\text{R},n};e_2,f_2),
\end{align}
where both $e_1$ and $f_1$ are the hyperparameters for $\boldsymbol{\vartheta}_\text{T}$, and both $e_2$ and $f_2$ are the hyperparameters for $\boldsymbol{\vartheta}_\text{R}$. Usually, we can choose the following values $a=b=c=d=e_1=f_1=e_2=f_2=10^{-2}$ as the hyperparameters. As shown in~\cite{Tipping2001}, the small values for hyperparameters are chosen and not sensitive to specific values~\cite{6194350}.

The off-grid parameter $\boldsymbol{\nu}$ follows a uniform prior distribution, and the distribution of the $u$-th entry $\nu_u$ can be expressed as
\begin{align}
	p(\nu_u;\delta)=\mathcal{U}_{\nu_u}\left(\left[-\frac{1}{2}\delta,\frac{1}{2}\delta\right]\right),
\end{align}
where we have 
\begin{align}
	\mathcal{U}_{x}\left(\left[a,b\right]\right)\triangleq\begin{cases}
		\frac{1}{b-a}, &a\leq x\leq b\\
		0,&\text{otherwise}
	\end{cases}.
\end{align}

The relationships between parameters are shown in Fig.~\ref{Bayesian}. To estimate the DOAs, we can formulate the following problem to maximize the posterior probability with the received signal 
\begin{align}
	\hat{\mathfrak{X}}=\arg\max_{\mathfrak{X}}p(\mathfrak{X}|\boldsymbol{R}),
\end{align}
where we use a set $\mathfrak{X}\triangleq  \left\{ \boldsymbol{X},\boldsymbol{\nu},\boldsymbol{c}_\text{T},\boldsymbol{c}_\text{R}, \sigma^2_n,\boldsymbol{\beta}\right\}$ to contain all the unknown parameters.
However, the problem of posterior probability cannot be solved directly, so an EM method is adopted to realize the sparse Bayesian learning. 

To obtain the posterior distribution of $\boldsymbol{X}$, we first calculate the joint distribution for all the parameters 
\begin{align}
	p(\boldsymbol{R},\mathfrak{X}) &=p(\boldsymbol{R}|\mathfrak{X})p(\boldsymbol{X}|\boldsymbol{\beta}) p(\boldsymbol{c}_\text{T}|\boldsymbol{\vartheta}_\text{T})p(\boldsymbol{c}_\text{R}|\boldsymbol{\vartheta}_\text{R})p( \alpha_n)\notag\\
	&\qquad p(\boldsymbol{\beta})p(\boldsymbol{\vartheta}_\text{T})p(\boldsymbol{\vartheta}_\text{R})p(\boldsymbol{\nu}).
\end{align}
Therefore, with the parameters $\alpha_n$, $\boldsymbol{\beta}$, $\boldsymbol{\vartheta}_\text{T}$, $\boldsymbol{\vartheta}_\text{R}$, $\boldsymbol{\nu}$, $\boldsymbol{c}_\text{T}$ and $\boldsymbol{c}_\text{R}$, the posterior for $\boldsymbol{X}$ can  be obtained as
\begin{align}
	&p(\boldsymbol{X}|\boldsymbol{R},\boldsymbol{\nu},\boldsymbol{c}_\text{T},\boldsymbol{c}_\text{R}, \alpha_n,\boldsymbol{\beta},\boldsymbol{\vartheta}_\text{T},\boldsymbol{\vartheta}_\text{R})\notag\\
	&\qquad  =\frac{p(\boldsymbol{R},\mathfrak{X})}{p(\boldsymbol{R},\boldsymbol{\nu},\boldsymbol{c}_\text{T},\boldsymbol{c}_\text{R}, \alpha_n,\boldsymbol{\beta},\boldsymbol{\vartheta}_\text{T},\boldsymbol{\vartheta}_\text{R})}\notag \\
	&\qquad = \frac{p(\boldsymbol{R}|\mathfrak{X})p(\boldsymbol{X}|\boldsymbol{\beta})}{p(\boldsymbol{R}|\boldsymbol{\nu},\boldsymbol{c}_\text{T},\boldsymbol{c}_\text{R}, \alpha_n,\boldsymbol{\beta},\boldsymbol{\vartheta}_\text{T},\boldsymbol{\vartheta}_\text{R})},\label{X}
\end{align}
where $p(\boldsymbol{R}|\mathfrak{X})$ and $(\boldsymbol{X}|\boldsymbol{\beta})$ can be calculated as
\begin{align}
	p(\boldsymbol{R}|\mathfrak{X}) &= \prod^{P-1}_{p=0}\mathcal{CN}(\boldsymbol{r}_p|\boldsymbol{\Upsilon}(\boldsymbol{\nu})(\boldsymbol{x}_p\otimes \boldsymbol{c}),\alpha_n^{-1}\boldsymbol{I}_{MN})\notag \\
	& = \prod^{P-1}_{p=0}\frac{\alpha_n^{MN}}{\pi^{MN}}e^{-\alpha_n\|\boldsymbol{r}_p-\boldsymbol{\Upsilon}(\boldsymbol{\nu})(\boldsymbol{x}_p\otimes \boldsymbol{c})\|^2_2},\\
p(\boldsymbol{X}|\boldsymbol{\beta})&=\prod_{p=0}^{P-1}\mathcal{CN}(\boldsymbol{x}_p|\boldsymbol{0}_{U\times 1}, \operatorname{diag}\{\boldsymbol{\beta}\}^{-1})\notag\\ 
& = \prod_{p=0}^{P-1} \left(\prod^{U-1}_{u=0}\beta_u\right) \frac{1}{\pi^U}e^{-\boldsymbol{x}^\text{H}_p\operatorname{diag}\{\boldsymbol{\beta}\}\boldsymbol{x}_p}. 
\end{align}
Since the denominator in (\ref{X}) is not a function of $\boldsymbol{X}$, the posterior distribution of $\boldsymbol{X}$ can be simplified as
\begin{align}
	p(\boldsymbol{X}|\boldsymbol{R},\boldsymbol{\nu},\boldsymbol{c}_\text{T},\boldsymbol{c}_\text{R}, \alpha_n,\boldsymbol{\beta},\boldsymbol{\vartheta}_\text{T},\boldsymbol{\vartheta}_\text{R})\propto p(\boldsymbol{R}|\mathfrak{X})p(\boldsymbol{X}|\boldsymbol{\beta}).
\end{align}
Both $ p(\boldsymbol{R}|\mathfrak{X})$ and $p(\boldsymbol{X}|\boldsymbol{\beta})$ are Gaussian functions, so the posterior for $\boldsymbol{X}$ can be also expressed as a Gaussian function
\begin{align}
	&p(\boldsymbol{X}|\boldsymbol{R},\boldsymbol{\nu},\boldsymbol{c}_\text{T},\boldsymbol{c}_\text{R}, \alpha_n,\boldsymbol{\beta},\boldsymbol{\vartheta}_\text{T},\boldsymbol{\vartheta}_\text{R}) \propto p(\boldsymbol{R}|\mathfrak{X})p(\boldsymbol{X}|\boldsymbol{\beta})\notag\\
	& \qquad \propto \prod^{P-1}_{p=0} e^{-\alpha_n\|\boldsymbol{r}_p-\boldsymbol{\Upsilon}(\boldsymbol{\nu})(\boldsymbol{I}_U\otimes \boldsymbol{c})\boldsymbol{x}_p\|^2_2-\boldsymbol{x}^\text{H}_p\operatorname{diag}\{\boldsymbol{\beta}\}\boldsymbol{x}_p}\notag \\
	&\qquad\triangleq \prod^{P-1}_{p=0}\mathcal{CN}(\boldsymbol{x}_p|\boldsymbol{\mu}_p,\boldsymbol{\Sigma}_\text{X}),
\end{align}
where the mean $\boldsymbol{\mu}_p$ and covariance matrix $\boldsymbol{\Sigma}_\text{X}$ are
\begin{align}
	\boldsymbol{\mu}_p&=\alpha_n \boldsymbol{\Sigma}_\text{X}(\boldsymbol{I}_U\otimes \boldsymbol{c})^\text{H} \boldsymbol{\Upsilon}^\text{H}(\boldsymbol{\nu})\boldsymbol{r}_p,\label{x}\\
	\boldsymbol{\Sigma}_\text{X}&=\left[\alpha_n(\boldsymbol{I}_{U}\otimes \boldsymbol{c})^\text{H} \boldsymbol{\Upsilon}^\text{H}(\boldsymbol{\nu})
	\boldsymbol{\Upsilon}(\boldsymbol{\nu})(\boldsymbol{I}_{U}
	\otimes \boldsymbol{c}) 
	+ \operatorname{diag}\{\boldsymbol{\beta}\}\right]^{-1}\label{sigma}.
\end{align}
and we use $\mu_{p,u}$ to denote the $u$-th entry of $\boldsymbol{\mu}_p$.

To calculate $\boldsymbol{\Sigma}_\text{X}$ and $\boldsymbol{\mu}_p$, we need to estimate the mutual coupling vectors $\boldsymbol{c}_\text{T}$ and $\boldsymbol{c}_\text{R}$, the off-grid parameter $\boldsymbol{\nu}$, and the \emph{precisions} $\alpha_n$ and $\boldsymbol{\beta}$. We can use the maximum posterior probability (MAP) method to maximize $p(\boldsymbol{\nu},\boldsymbol{c}_\text{T},\boldsymbol{c}_\text{R}, \alpha_n,\boldsymbol{\beta},\boldsymbol{\vartheta}_\text{T},\boldsymbol{\vartheta}_\text{R}|\boldsymbol{R})$. We have
\begin{align}
	&p(\boldsymbol{\nu},\boldsymbol{c}_\text{T},\boldsymbol{c}_\text{R}, \alpha_n,\boldsymbol{\beta},\boldsymbol{\vartheta}_\text{T},\boldsymbol{\vartheta}_\text{R}|\boldsymbol{R})p(\boldsymbol{R})\notag\\
	&\qquad =p(\boldsymbol{\nu},\boldsymbol{c}_\text{T},\boldsymbol{c}_\text{R}, \alpha_n,\boldsymbol{\beta},\boldsymbol{\vartheta}_\text{T},\boldsymbol{\vartheta}_\text{R},\boldsymbol{R}),
	\end{align}
so maximizing $p(\boldsymbol{\nu},\boldsymbol{c}_\text{T},\boldsymbol{c}_\text{R}, \alpha_n,\boldsymbol{\beta},\boldsymbol{\vartheta}_\text{T},\boldsymbol{\vartheta}_\text{R}|\boldsymbol{R})$ is equivalent to maximizing $p(\boldsymbol{\nu},\boldsymbol{c}_\text{T},\boldsymbol{c}_\text{R}, \alpha_n,\boldsymbol{\beta},\boldsymbol{\vartheta}_\text{T},\boldsymbol{\vartheta}_\text{R},\boldsymbol{R})$. The EM method can be used to solve the MAP estimation by treating $\boldsymbol{X}$ as a hidden variable. Before estimating the parameters, we will first obtain the likelihood function under the expectation with respect to the posterior of $\boldsymbol{X}$
\begin{align}
	&\mathcal{L}(\boldsymbol{\nu},\boldsymbol{c}_\text{T},\boldsymbol{c}_\text{R}, \alpha_n,\boldsymbol{\beta},\boldsymbol{\vartheta}_\text{T},\boldsymbol{\vartheta}_\text{R})
	\notag \\
	& \triangleq \mathcal{E}_{\boldsymbol{X}|\boldsymbol{R},\boldsymbol{\nu},\boldsymbol{c}_\text{T},\boldsymbol{c}_\text{R}, \alpha_n,\boldsymbol{\beta},\boldsymbol{\vartheta}_\text{T},\boldsymbol{\vartheta}_\text{R}}\left\{\ln p(\mathfrak{X},\boldsymbol{\vartheta}_\text{T},\boldsymbol{\vartheta}_\text{R},\boldsymbol{R}) \right\}. 
	\end{align}
To simplify the notation, we just use $\mathcal{E}\{\cdot\}$ to represent $\mathcal{E}_{\boldsymbol{X}|\boldsymbol{R},\boldsymbol{\nu},\boldsymbol{c}_\text{T},\boldsymbol{c}_\text{R}, \alpha_n,\boldsymbol{\beta},\boldsymbol{\vartheta}_\text{T},\boldsymbol{\vartheta}_\text{R}}\{\cdot\}$, so the likelihood function can be simplified as
	\begin{align}
	& \mathcal{L}(\boldsymbol{\nu},\boldsymbol{c}_\text{T},\boldsymbol{c}_\text{R}, \alpha_n,\boldsymbol{\beta},\boldsymbol{\vartheta}_\text{T},\boldsymbol{\vartheta}_\text{R})\notag\\
	&\qquad = \mathcal{E}\big\{\ln  p(\boldsymbol{R}|\mathfrak{X})p(\boldsymbol{X}|\boldsymbol{\beta}) p(\boldsymbol{c}_\text{T}|\boldsymbol{\vartheta}_\text{T})p(\boldsymbol{c}_\text{R}|\boldsymbol{\vartheta}_\text{R})p( \alpha_n)\notag\\
	&\qquad\qquad\quad p(\boldsymbol{\beta})p(\boldsymbol{\vartheta}_\text{T})p(\boldsymbol{\vartheta}_\text{R})p(\boldsymbol{\nu})\big\}. 
\end{align}

In the following contents, we will give the expressions for all the remaining unknown parameters.

\begin{enumerate}
	\item  For the mutual coupling vector $\boldsymbol{c}_\text{T}$, ignoring terms independent thereof, we can obtain the following likelihood function  
\begin{align}
	\mathcal{L}&(\boldsymbol{\boldsymbol{c}_\text{T}})  = \mathcal{E}\left\{\ln  p(\boldsymbol{R}|\boldsymbol{X},\boldsymbol{\nu},\boldsymbol{c}_\text{T},\boldsymbol{c}_\text{R}, \alpha_n) p(\boldsymbol{c}_\text{T}|\boldsymbol{\vartheta}_\text{T})\right\}\notag\\
	&= \mathcal{E}\left\{\ln   \prod^{P-1}_{p=0}\mathcal{CN}(\boldsymbol{r}_p|\boldsymbol{\Upsilon}(\boldsymbol{\nu})(\boldsymbol{x}_p\otimes \boldsymbol{c}),\alpha^{-1}_n\boldsymbol{I}_{MN})\right\}\notag\\
	&\qquad + \ln
	 \prod^{M-1}_{m=0}\mathcal{CN}(c_{\text{T},m}|0,\vartheta^{-1}_{\text{T},m})\notag\\
	 	 &\propto 
	 -\alpha_n P \operatorname{Tr}\left\{
	 (\boldsymbol{I}_U\otimes \boldsymbol{c})^\text{H}
	\boldsymbol{\Upsilon}^\text{H}(\boldsymbol{\nu})
	 \boldsymbol{\Upsilon}(\boldsymbol{\nu})(\boldsymbol{I}_U\otimes \boldsymbol{c})\boldsymbol{\Sigma}_\text{X}\right\} \notag\\
	 &\qquad -\sum^{P-1}_{p=0} \alpha_n  \|\boldsymbol{r}_p-\boldsymbol{\Upsilon}(\boldsymbol{\nu})(\boldsymbol{\mu}_p\otimes \boldsymbol{c})\|^2_2\notag\\
	 & \qquad - \sum^{M-1}_{m=0}  \vartheta_{\text{T},m} |c_{\text{T},m}|^2.
	 \end{align}
In Appendix~\ref{ct}	, the details about calculating $\frac{\partial \mathcal{L}(\boldsymbol{c}_\text{T})}{\partial \boldsymbol{c}_\text{T}}$ are given. By setting $\frac{\partial \mathcal{L}(\boldsymbol{c}_\text{T})}{\partial \boldsymbol{c}_\text{T}}=\boldsymbol{0}$,  $\boldsymbol{c}_\text{T}$ can be obtained as 
\begin{align}\label{ctE}
	\boldsymbol{c}_\text{T}=\boldsymbol{H}^{-1}_\text{T}\boldsymbol{z}_\text{T},
\end{align}
where
\begin{align}
	\boldsymbol{H}_\text{T}& =  \sum^{P-1}_{p=0} \alpha_n  
	 \boldsymbol{T}^\text{H}_\text{T}
		\boldsymbol{\Upsilon}^\text{H}(\boldsymbol{\nu}) 
		\boldsymbol{\Upsilon}(\boldsymbol{\nu})(\boldsymbol{\mu}_p\otimes \boldsymbol{c}_\text{R}\otimes \boldsymbol{I}_M)\notag\\
		&  + \alpha_n P	 
	\boldsymbol{G}_\text{T}^\text{H}\left(\sum_{p=0}^{U-1}\sum_{k=0}^{U-1}
	 \boldsymbol{\Upsilon}^\text{H}_p(\boldsymbol{\nu})\boldsymbol{\Upsilon}_k(\boldsymbol{\nu})
	 \Sigma_{\text{X},k,p}
	 \right)^\text{H}\notag\\
	 &\quad (\boldsymbol{c}_\text{R}\otimes\boldsymbol{I}_M) 
		+\operatorname{diag}\{\boldsymbol{\vartheta_\text{T}}\},
		\end{align}
and
\begin{align}
		\boldsymbol{z}_\text{T} &=  \sum^{P-1}_{p=0} \alpha_n  
		\boldsymbol{T}_\text{T}^\text{H}
		\boldsymbol{\Upsilon}^\text{H}(\boldsymbol{\nu}) 
		\boldsymbol{r}_p, \\
		\boldsymbol{T}_\text{T}& \triangleq \begin{bmatrix}
		\boldsymbol{\mu}_p\otimes \boldsymbol{c}_\text{R}\otimes \boldsymbol{e}^M_0 ,
			\dots,\boldsymbol{\mu}_p\otimes \boldsymbol{c}_\text{R}\otimes \boldsymbol{e}^M_{M-1} 
		\end{bmatrix},
		\\
		\boldsymbol{G}_\text{T}& \triangleq \begin{bmatrix}
	 	\boldsymbol{c}_\text{R}\otimes \boldsymbol{e}^M_0,\boldsymbol{c}_\text{R}\otimes \boldsymbol{e}^M_1,\dots,\boldsymbol{c}_\text{R}\otimes \boldsymbol{e}^M_{M-1}
	 \end{bmatrix}.
		\end{align}

\item For the mutual coupling vector $\boldsymbol{c}_\text{R}$, using the same method with $\boldsymbol{c}_\text{T}$, we can obtain
\begin{align}\label{crE}
\boldsymbol{c}_\text{R}= \boldsymbol{H}^{-1}_\text{R}\boldsymbol{z}_\text{R},
\end{align}	 
where
\begin{align}
	\boldsymbol{H}_\text{R} =  & \sum^{P-1}_{p=0} \alpha_n  
	 \boldsymbol{T}^\text{H}_\text{R}
		\boldsymbol{\Upsilon}^\text{H}(\boldsymbol{\nu}) 
		\boldsymbol{\Upsilon}(\boldsymbol{\nu})(\boldsymbol{\mu}_p\otimes \boldsymbol{I}_N\otimes \boldsymbol{c}_\text{T}) 
\notag\\
&+\alpha_n P	 
	\boldsymbol{G}_\text{R}^\text{H}\left(\sum_{p=0}^{U-1}\sum_{k=0}^{U-1}
	 \boldsymbol{\Upsilon}^\text{H}_p(\boldsymbol{\nu})\boldsymbol{\Upsilon}_k(\boldsymbol{\nu})
	 \Sigma_{\text{X},k,p}
	 \right)^\text{H}\notag\\
	 &(\boldsymbol{I}_N\otimes\boldsymbol{c}_\text{T})  +\operatorname{diag}\{\boldsymbol{\vartheta_\text{T}}\}, 
		\end{align}
and
		\begin{align}
		\boldsymbol{z}_\text{R} &=  \sum^{P-1}_{p=0} \alpha_n  
		\boldsymbol{T}_\text{R}^\text{H}
		\boldsymbol{\Upsilon}^\text{H}(\boldsymbol{\nu}) 
		\boldsymbol{r}_p, \\
		\boldsymbol{T}_\text{R}& \triangleq \begin{bmatrix}
		\boldsymbol{\mu}_p\otimes  \boldsymbol{e}^N_0 \otimes \boldsymbol{c}_\text{T},
			\dots,\boldsymbol{\mu}_p\otimes \boldsymbol{e}^N_{N-1} \otimes \boldsymbol{c}_\text{T}
		\end{bmatrix},
		\\
		\boldsymbol{G}_\text{R}& \triangleq \begin{bmatrix}
	 	\boldsymbol{e}^N_0\otimes \boldsymbol{c}_\text{T},  \boldsymbol{e}^N_1\boldsymbol{c}_\text{T},\dots,  \boldsymbol{e}^N_{N-1}\boldsymbol{c}_\text{T}
	 \end{bmatrix}.
		\end{align}

\item For the \emph{precision} $\boldsymbol{\beta}$ of scattering coefficients, ignoring terms independent thereof, we can obtain the likelihood function
\begin{align}
	&\mathcal{L}(\boldsymbol{\beta})  = \mathcal{E} \left\{\ln  p(\boldsymbol{X}|\boldsymbol{\beta}) p(\boldsymbol{\beta}) \right\}\notag\\
	&= \mathcal{E}\left\{\ln  \prod_{p=0}^{P-1}\mathcal{CN}(\boldsymbol{x}_p|\boldsymbol{0}_{U\times 1}, \boldsymbol{\Lambda}_x)\right\} +\ln
	 \prod^{U-1}_{u=0}\mathfrak{G}(\beta_u;c,d). 
\end{align}

By setting $\frac{\partial \mathcal{L}(\boldsymbol{\beta})}{\partial \boldsymbol{\beta}}=0$, the $u$-th entry of $\boldsymbol{\beta}$ can be obtained as 
\begin{align}\label{beta}
	\beta_u =\frac{P+c-1}{d + P\Sigma_{\text{X},u,u} + \sum_{p=0}^{P-1} |\mu_{u,p}|^2}. 
\end{align}

\item For the \emph{precision} $\alpha_n$ of noise, ignoring terms independent thereof, we can obtain the likelihood function
\begin{align}
	&\mathcal{L}(\alpha_n)= \mathcal{E}\left\{\ln  p(\boldsymbol{R}|\boldsymbol{X},\boldsymbol{\nu},\boldsymbol{c}_\text{T},\boldsymbol{c}_\text{R}, \alpha_n)p(\alpha_n) \right\}\notag\\
	& = \mathcal{E}\left\{\ln  
	\prod_{p=0}^{P-1}\mathcal{CN}\left(\boldsymbol{r}_p|\boldsymbol{\Upsilon}(\boldsymbol{\nu})(\boldsymbol{x}_p\otimes \boldsymbol{c}_\text{R}\otimes \boldsymbol{c}_\text{T}), \sigma^2_n\boldsymbol{I}\right)
	 \right\}\notag\\
	 &\qquad +\ln
	\mathfrak{G}(\alpha_n;a,b).
\end{align}
By setting $\frac{\partial \mathcal{L}(\alpha_n)}{\partial \alpha_n}=0$, we can obtain
\begin{align}\label{n}
\alpha_n&=\frac{MNP+a-1}{P\mathfrak{N}_1 + \mathfrak{N}_2
	  +b},
\end{align}
where 
\begin{align}
	\mathfrak{N}_1&\triangleq \operatorname{Tr}\{(\boldsymbol{I}_U\otimes \boldsymbol{c})^\text{H}\boldsymbol{\Upsilon}^\text{H}(\boldsymbol{\nu})
	\boldsymbol{\Upsilon}(\boldsymbol{\nu}) 
	(\boldsymbol{I}_U\otimes \boldsymbol{c})\boldsymbol{\Sigma}_\text{X}\},\\
	\mathfrak{N}_2&\triangleq \|\boldsymbol{R}-\boldsymbol{\Upsilon}(\boldsymbol{\nu})(\boldsymbol{\mu}\otimes \boldsymbol{c})\|^2_F,\\
	\boldsymbol{\mu}&\triangleq\begin{bmatrix}
	\boldsymbol{\mu}_0,\boldsymbol{\mu}_1,\dots,\boldsymbol{\mu}_{P-1}
\end{bmatrix}.
\end{align}

\item For the \emph{precision} $\boldsymbol{\vartheta}_\text{T}$ of mutual coupling vector, ignoring terms independent thereof, we can obtain the likelihood function 
\begin{align}
	\mathcal{L}(\boldsymbol{\vartheta}_\text{T}) &= \mathcal{E}\left\{\ln p(\boldsymbol{c}_\text{T}|\boldsymbol{\vartheta}_\text{T}) p(\boldsymbol{\vartheta}_\text{T}) \right\}\notag\\
	& = \mathcal{E}\left\{\ln  
	\prod^{M-1}_{m=0}\mathcal{CN}(c_{\text{T},m}|0,\sigma^2_{\text{T},m})\right\}\notag \\
	&\qquad+\ln
	\prod^{M-1}_{m=0}\mathfrak{G}(\vartheta_{\text{T},m};e_1,f_1).  
\end{align}
By setting $\frac{\partial \mathcal{L}(\boldsymbol{\vartheta}_\text{T})}{\partial \boldsymbol{\vartheta}_\text{T}}=\boldsymbol{0}$, we can obtain the $m$-th entry of $\boldsymbol{\vartheta}_\text{T}$ as 
\begin{align}\label{varT}
	 \vartheta_{\text{T},m}= \frac{e_1}{f_1+c_{\text{T},m}^\text{H}c_{\text{T},m}}.
\end{align}

\item For the \emph{precision} $\boldsymbol{\vartheta}_\text{R}$ of mutual coupling vector, using the same method, we can obtain the $n$-th entry of $\boldsymbol{\vartheta}_\text{R}$ as 
\begin{align}\label{varR}
	 \vartheta_{\text{R},n}= \frac{e_2}{f_2+c_{\text{R},n}^\text{H}c_{\text{R},n}}.
\end{align}

\item For off-grid $\boldsymbol{\nu}$, ignoring terms independent thereof, we can obtain the likelihood function
\begin{align} 
	&\mathcal{L}(\boldsymbol{\nu})  = \mathcal{E} \left\{\ln  p(\boldsymbol{R}|\boldsymbol{X},\boldsymbol{\nu},\boldsymbol{c}_\text{T},\boldsymbol{c}_\text{R}, \alpha_n) p(\boldsymbol{\nu})\right\}.
\end{align} 

By setting $\frac{\partial \mathcal{L}(\boldsymbol{\nu})}{\partial \boldsymbol{\nu}}=0$, we can obtain
\begin{align}\label{nu}
	\boldsymbol{\nu}=\boldsymbol{H}^{-1}\boldsymbol{z},
\end{align}
where the entry of the $u$-th row and $m$-column in $\boldsymbol{H}\in\mathbb{R}^{U\times U}$ is
\begin{align}
	H_{u,m}=\mathcal{R}\left\{\left(P\Sigma_{\text{X},u,m}+\sum^{P-1}_{p=0}
	 \mu^\text{H}_{p,m} \mu_{p,u}\right)\boldsymbol{c}^\text{H} \boldsymbol{\Xi}^\text{H}_m\boldsymbol{\Xi}_u\boldsymbol{c}
  \right\},
\end{align}
and the $u$-th entry of $\boldsymbol{z}\in\mathcal{R}^{U\times 1}$ is
\begin{align}
	z_u&=\sum^{P-1}_{p=0}\mathcal{R}\left\{\left[\boldsymbol{r}_p
- \boldsymbol{\Psi}(\boldsymbol{\mu}_p\otimes \boldsymbol{c})
\right]^\text{H}\boldsymbol{\Xi}_u \mu_{u,p}\boldsymbol{c}\right\}\notag\\
&\qquad -\sum_{m=0}^{U-1}\mathcal{R}\left\{P\Sigma_{\text{X},u,m}\boldsymbol{c}^\text{H} \boldsymbol{\Psi}^\text{H}_m\boldsymbol{\Xi}_u\boldsymbol{c}\right\}.
\end{align}
The details to obtain $\boldsymbol{\nu}$ are given in Appendix~\ref{nuD}.

\end{enumerate} 
 
In Algorithm~\ref{alg1}, we show the details about the proposed method SBLMC to estimate the DOAs with unknown mutual coupling effect. In the proposed SBLMC algorithm, after the iterations, we can obtain the spatial spectrum $\boldsymbol{P}_{\text{X}}$ of the sparse matrix $\boldsymbol{X}$ from the received signal $\boldsymbol{R}$. Then, by searching all the values of  $\boldsymbol{P}_{\text{X}}$, the corresponding peak values can be found. By selecting positions of peak values corresponding to the $K$ maximum values, we can estimate the DOAs of targets, where we use $\boldsymbol{\zeta}+\boldsymbol{\nu}$ as the discretized angle vector.

\begin{algorithm}
	\caption{SBLMC algorithm to estimate the DOAs with unknown mutual coupling effect} \label{alg1}
	\begin{algorithmic}[1]
		\STATE  \emph{Input:} received signal $\boldsymbol{R}$, dictionary matrix $\boldsymbol{\Psi}$, the first order derivative of dictionary matrix $\boldsymbol{\Xi}$, the number of pulses $P$, the maximum of iteration $N_{\text{iter}}$, stop threshold $\lambda_{\text{th}}$.
		\STATE \emph{Initialization:} $\boldsymbol{c}_T=\boldsymbol{\vartheta}_T=[1,\boldsymbol{0}_{1\times (M-1)}]^T$, $\boldsymbol{c}_R=\boldsymbol{\vartheta}_R=[1,\boldsymbol{0}_{1\times (N-1)}]^T$, $\alpha_n=1$, the hyperparameters $a=b=c=d=e_1=f_1=e_2=f_2=10^{-2}$, $\boldsymbol{\nu}=\boldsymbol{0}_{U\times 1}$, $\boldsymbol{\beta}=\boldsymbol{1}_{U\times 1}$, $i_{\text{iter}}=1$, $\lambda = \|\boldsymbol{R}\|^2_F$.
		\WHILE{$i_{\text{iter}} \leq N_{\text{iter}}$ or $\lambda\leq \lambda_{\text{th}}$}
		\STATE \label{4}$\boldsymbol{\Upsilon}(\boldsymbol{\nu})\leftarrow \boldsymbol{\Psi}+\boldsymbol{\Xi}\left(\operatorname{diag}\left\{
	\boldsymbol{\nu} \right\}\otimes \boldsymbol{I}_{MN}\right)$.
		\STATE \label{5} Obtain $\boldsymbol{\mu}_p$ ($p=0,1,\dots,P-1$) and $\boldsymbol{\Sigma}_{\text{X}}$ from (\ref{x}) and (\ref{sigma}), respectively.
		\STATE \label{6} Obtain the spatial spectrum 
		\begin{align}
			P_\text{X}=\mathcal{R}\left\{\operatorname{diag}\{\boldsymbol{\Sigma}_{\text{X}}\}\right\} + \frac{1}{P}\sum^{P-1}_{p=0}|\boldsymbol{\mu}_p|^2,
		\end{align}
		where $|\boldsymbol{\mu}_p|\triangleq\begin{bmatrix}
			|\mu_{p,0}|,|\mu_{p,1}|,\dots,|\mu_{p,U-1}|
		\end{bmatrix}^T$.
		\STATE \label{7} $\boldsymbol{\beta}'\leftarrow \boldsymbol{\beta}$, and update $\boldsymbol{\beta}$ from (\ref{beta}).
		\STATE \label{8} Update $\boldsymbol{c}_T$ and $\boldsymbol{c}_R$ from (\ref{ctE}) and (\ref{crE}), respectively.
		\STATE \label{9} Update $\boldsymbol{\vartheta}_T$ and $\boldsymbol{\vartheta}_R$ from (\ref{varT}) and (\ref{varR}), respectively.
		\STATE \label{10} Estimate $\boldsymbol{\nu}$ from (\ref{nu}).
		\STATE \label{11} Update $\alpha_n$ from (\ref{n}).
		\IF{$i_{\text{iter}}>1$}
		\STATE $\lambda = \frac{\|\boldsymbol{\beta}-\boldsymbol{\beta}'\|_2}{\|\boldsymbol{\beta}'\|_2}$.
		\ENDIF
		\STATE $i_{\text{iter}}\leftarrow i_{\text{iter}}+1$.
		\ENDWHILE
		\STATE \emph{Output:} the spatial spectrum $P_{\text{X}}$, and the DOAs $(\boldsymbol{\zeta}+\boldsymbol{\nu})$ can be obtained from the positions of peak values in $P_{\text{X}}$.
	\end{algorithmic}
\end{algorithm}

\section{Simulation Results}\label{sec4}
\begin{table}[!t]
	\renewcommand{\arraystretch}{1.3}
	\caption{Simulation Parameters}
	\label{table1}
	\centering
	\begin{tabular}{cc}
		\hline
		\textbf{Parameter} & \textbf{Value}\\
		\hline
		The signal-to-noise ratio (SNR) of echo signal & $ 20 $ dB\\
		The number of pulses $P$ & $100$\\
		The number of transmitting antennas $M$ & $10$\\
		The number of receiving antennas $N$ & $5$\\
		The number of targets $K$& $3$\\
		The space between antennas $d_T=d_R$& $0.5$ wavelength\\
		The grid size $\delta$ & $\ang{2}$\\
		The detection DOA range & $\left[\ang{-80},\ang{80}\right]$\\
		The hyperparameters $a,b,c,d,e_1,f_1,e_2,f_2$ & $10^{-2}$\\
		The mutual coupling between adjacent antennas & $-5$ dB\\
	\hline
	\end{tabular}
\end{table}

\begin{figure}
	\centering
	\includegraphics[width=3.5in]{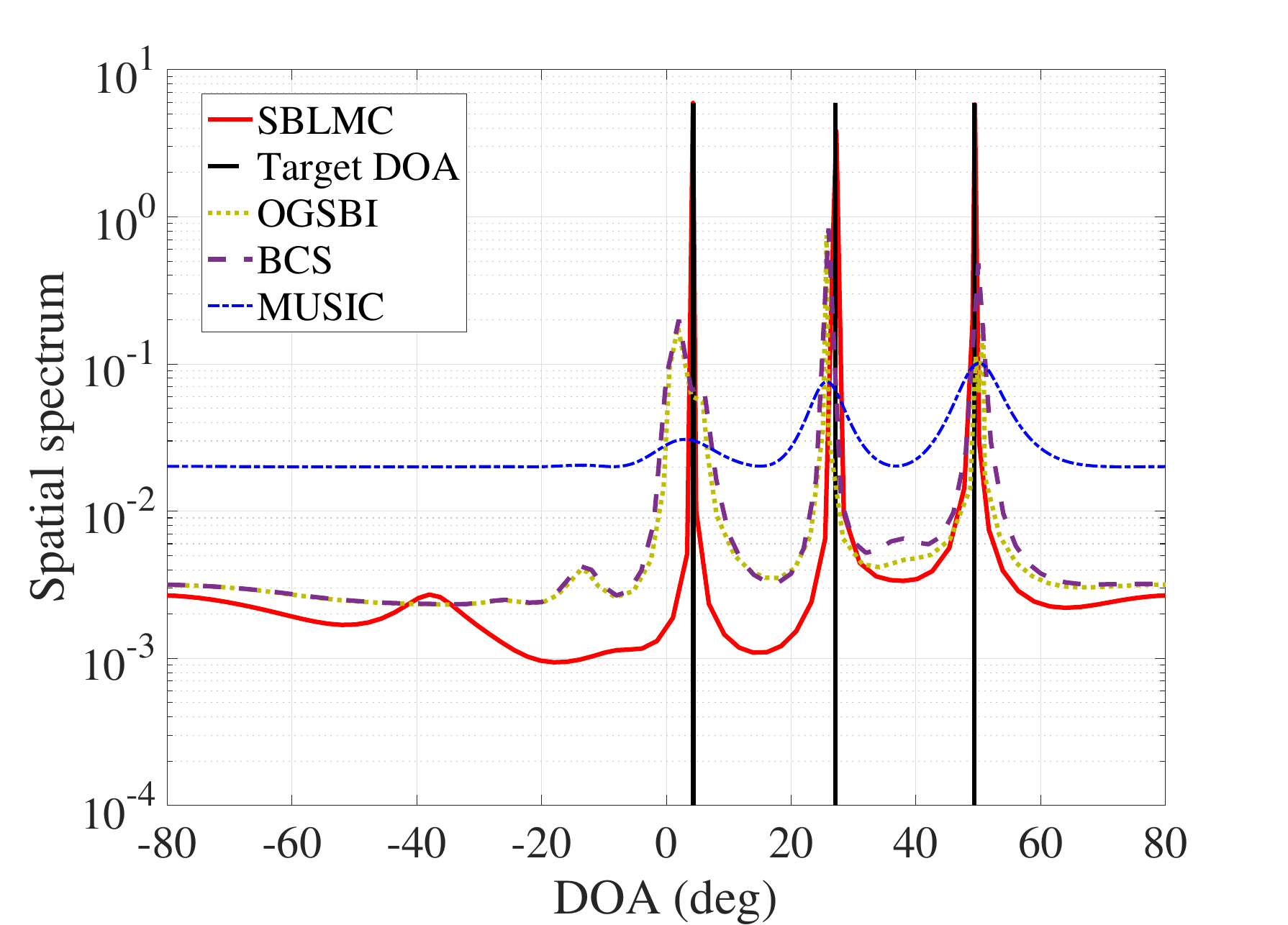}
	\caption{The spatial spectrum for DOA estimation.}
	\label{spectrum}
\end{figure}

\begin{figure}
	\centering
	\includegraphics[width=3.5in]{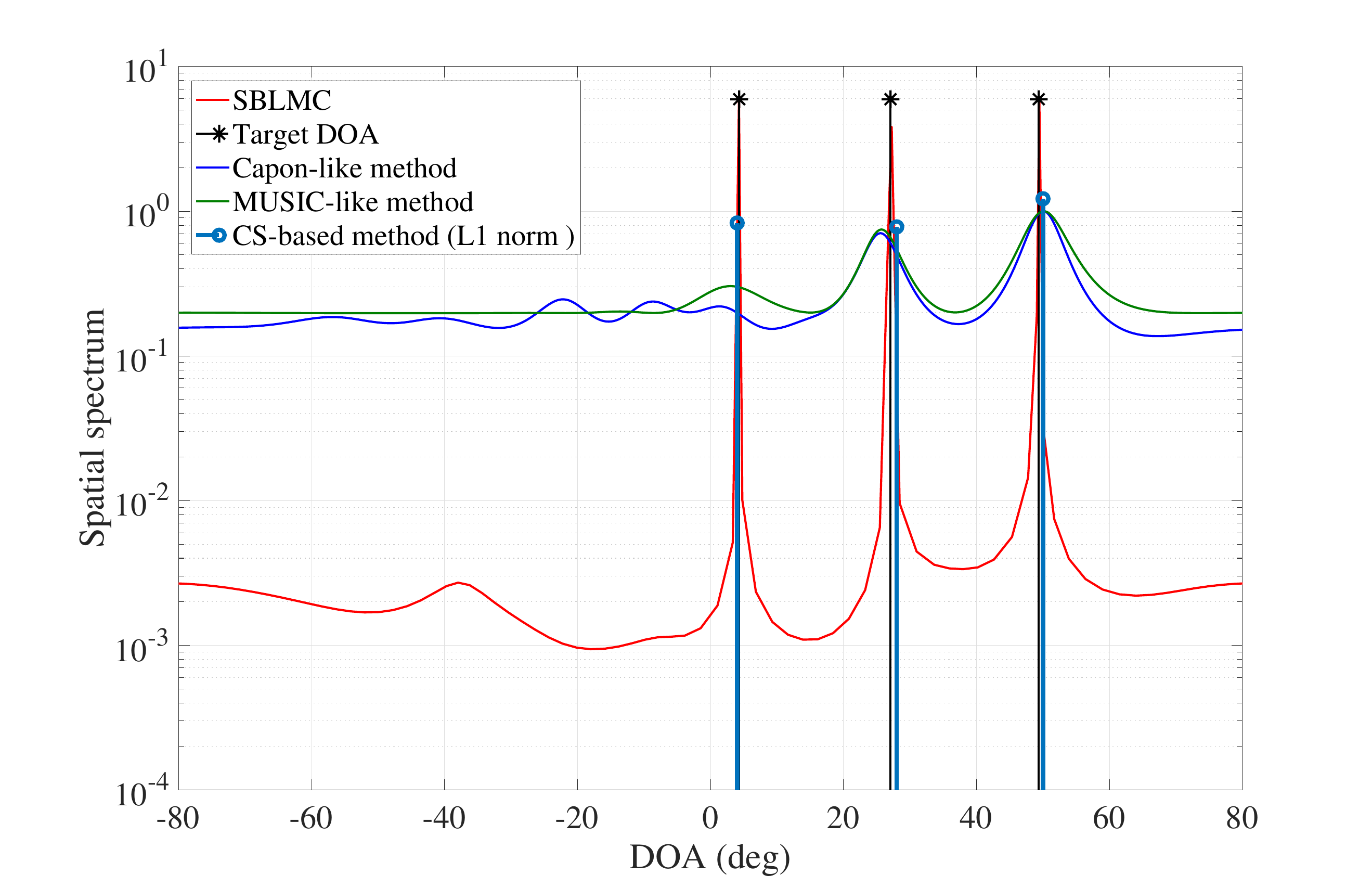}
	\caption{The spatial spectrum of the proposed method is compared with that of the present methods in DOA estimation.}
	\label{compare}
\end{figure}

\begin{figure}
	\centering
	\includegraphics[width=3.5in]{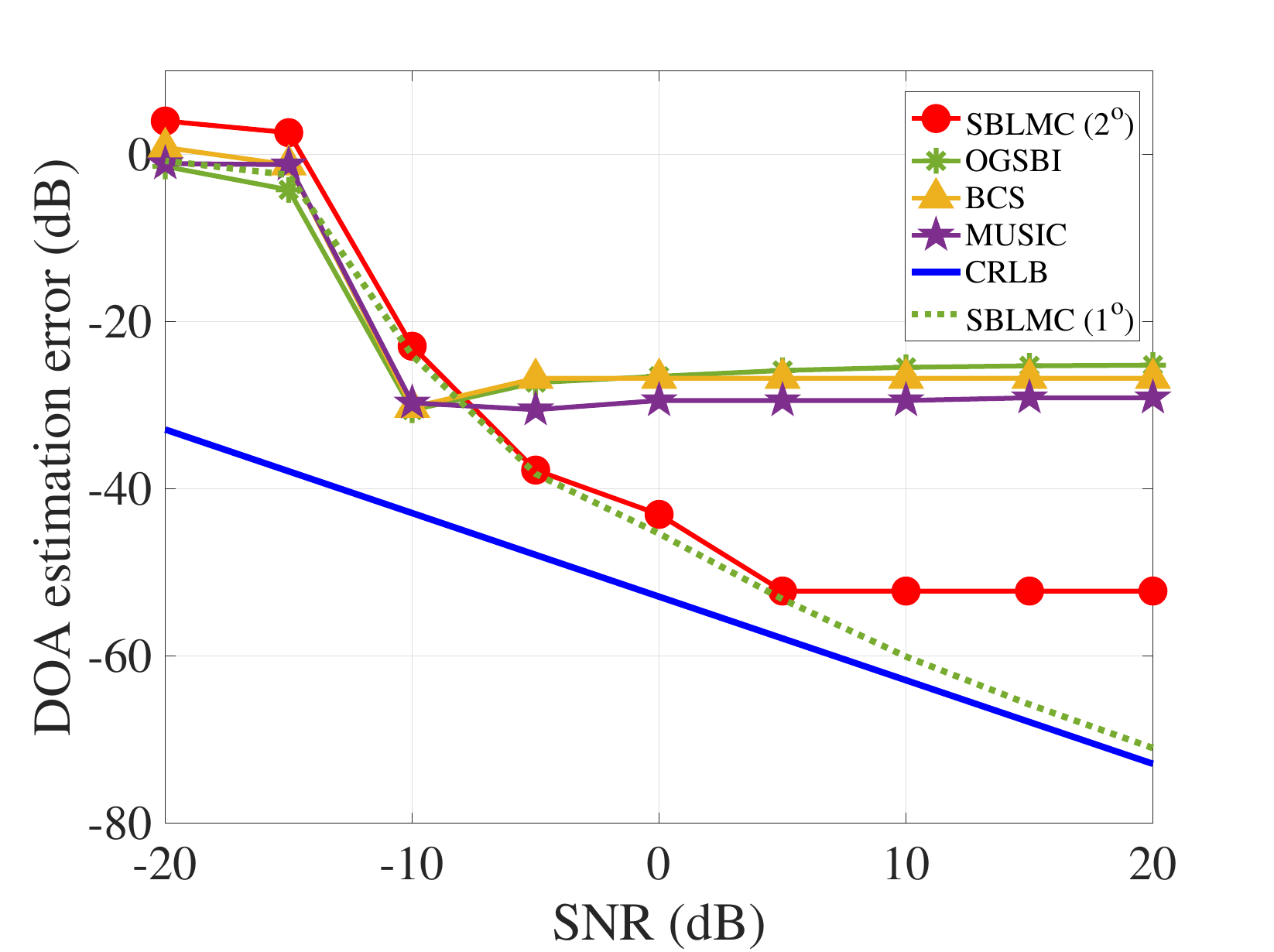}
	\caption{The DOA estimation performance with different SNRs.}
	\label{SNR}
\end{figure}

\begin{figure}
	\centering
	\includegraphics[width=3.5in]{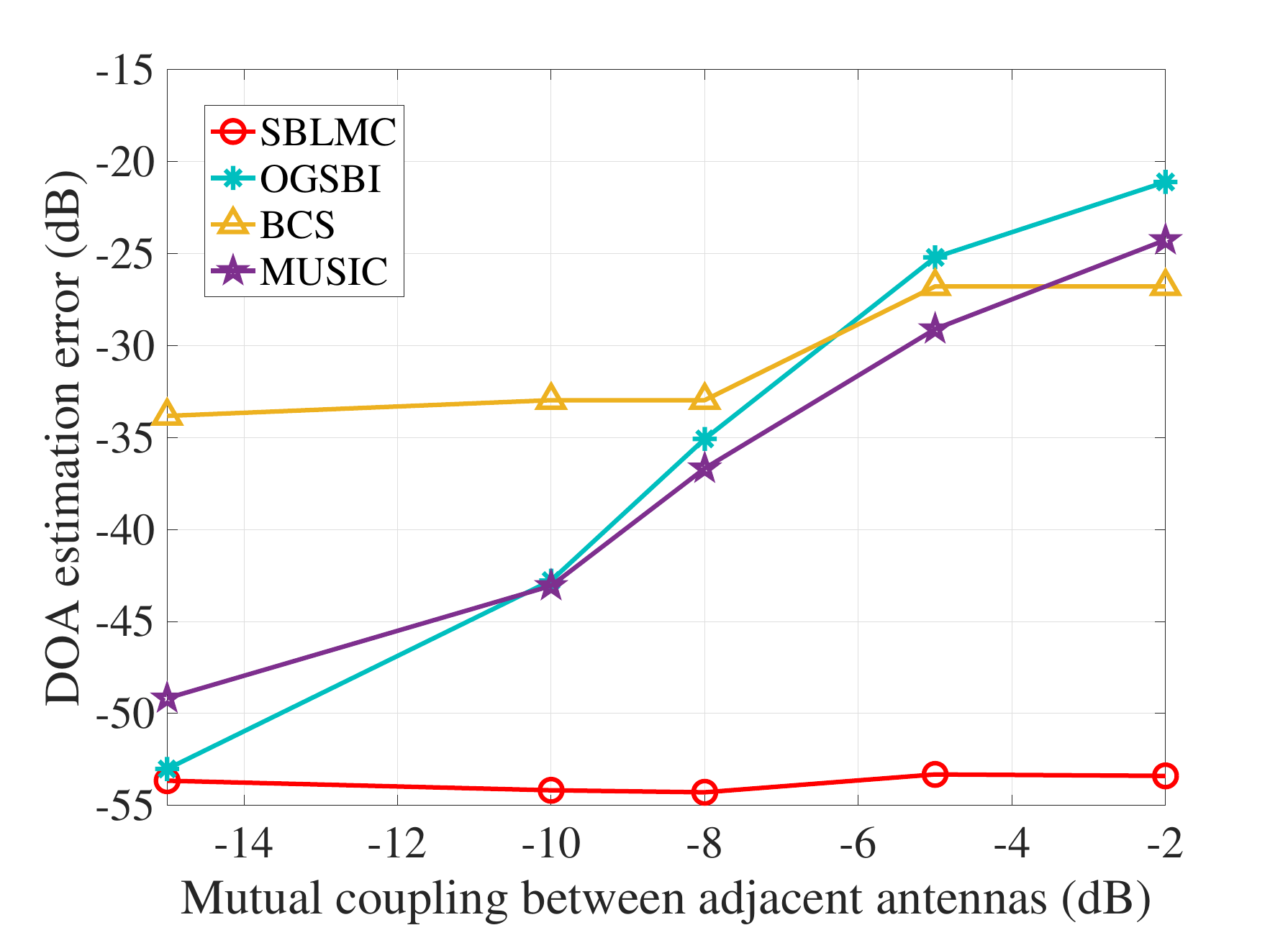}
	\caption{The DOA estimation performance with different mutual coupling effects.}
	\label{MC}
\end{figure}

\begin{figure}
	\centering
	\includegraphics[width=3.5in]{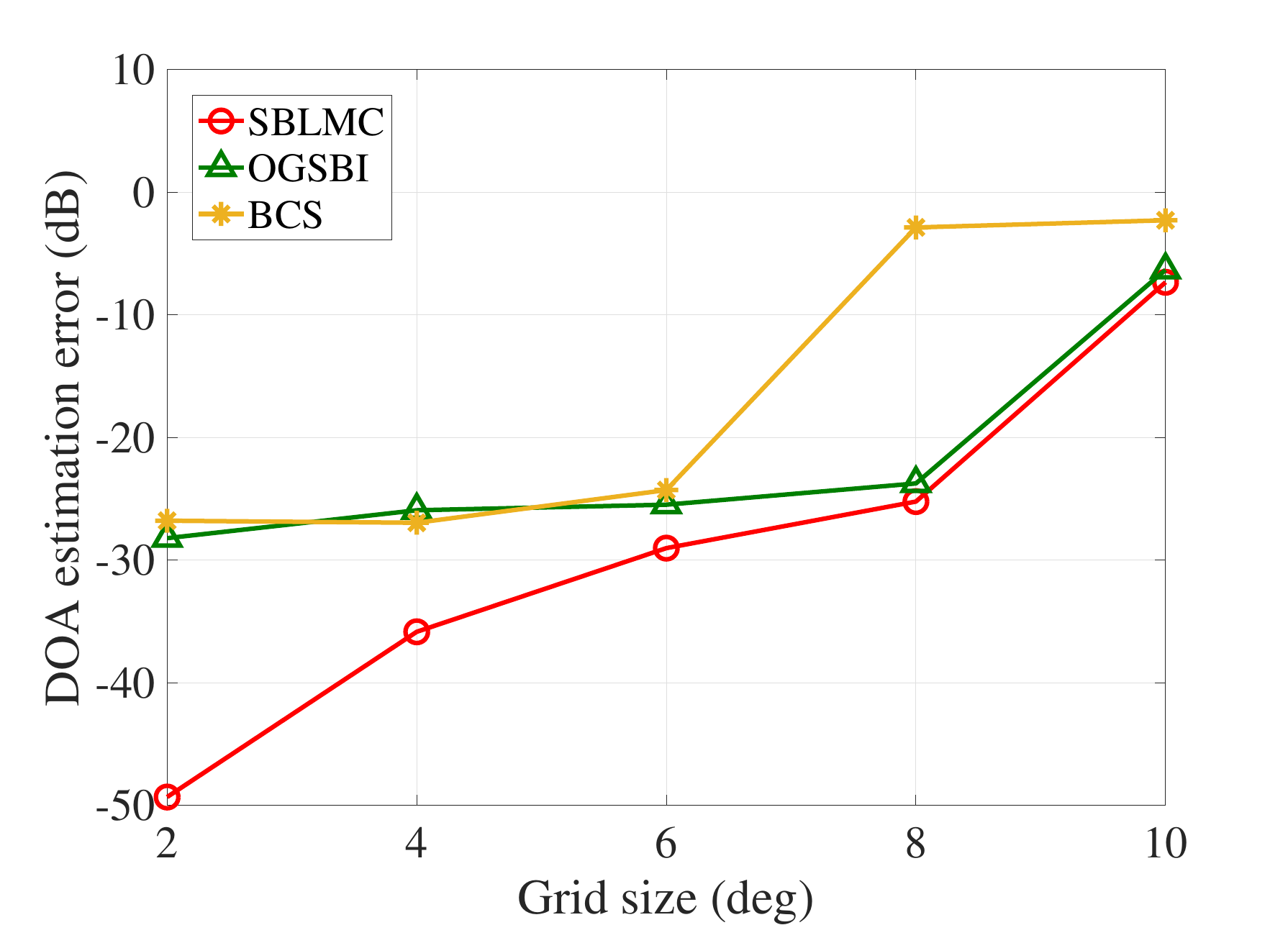}
	\caption{The DOA estimation performance with different grid sizes.}
	\label{grid}
\end{figure}

\begin{table}[!t]
	\renewcommand{\arraystretch}{1.3}
	\caption{Estimated DOAs}
	\label{table2}
	\centering
	\begin{tabular}{cccc}
		\hline
		\textbf{Methods} & \textbf{Target $1$} & \textbf{Target $2$} & \textbf{Target $3$}\\
		\hline
		\textbf{Target DOAs} &$\ang{4.3075}$& $\ang{27.0740}$& $\ang{49.3603}$\\
		\textbf{SBLMC}&$\ang{4.2746}$&$\ang{27.2441}$&\ang{49.4521}\\
		\textbf{OGSBI}& $\ang{1.8636}$ & $\ang{25.6868}$&
  $\ang{50.7775}$\\
		\textbf{BCS}& $\ang{2.0000}$ & $\ang{26.0000}$&
    $\ang{50.0000}$\\
		\textbf{MUSIC}&
		$\ang{2.9929}$&$\ang{25.7086}$&$\ang{50.1324}$\\
	\hline
	\end{tabular}
\end{table}

In this section, the simulation results about the proposed method for DOA estimation in the MIMO radar system are given, and the simulation parameters are given in Table~\ref{table1}. For the proposed SBLMC algorithm, the maximum iteration is $N_{\text{iter}}=10^3$ and the stop threshold is $\lambda=10^{-3}$. All experiments are carried out in Matlab R2017b on a PC with a 2.9 GHz Intel Core i5 and 8 GB of RAM. Matlab codes have been made available online at \url{https://sites.google.com/site/chenpengdsp/publications}.

First, we show the estimated spatial spectrum of $3$ targets. As shown in Fig.~\ref{spectrum}, $3$ present methods including off-grid sparse Bayesian inference (OGSBI)~\cite{yang2013}, Bayesian compressive sensing (BCS)~\cite{shihao2008} and MUSIC~\cite{schmidt1981}, have compared with the proposed SBLMC method. With the mutual coupling effect, the traditional MUSIC method cannot achieve better performance. Since the present Bayesian methods (OGSBI and BCS) have not considered the mutual coupling effect, the estimation performance cannot be further improved. The proposed SBLMC considers both off-grid and mutual coupling effects, can achieve the best spatial spectrum and improve the DOA estimation performance. In Table~\ref{table2}, we give the estimated DOAs with different methods. We use the following expression to measure the estimation performance
\begin{align}
	e\triangleq 10\log_{10}\|\hat{\boldsymbol{\theta}}-\boldsymbol{\theta}\|^2_2\quad(\text{dB}),
\end{align}
where $\hat{\boldsymbol{\theta}}$ denotes the estimated DOA vector and $\boldsymbol{\theta}$ is the target DOA vector. Both $\hat{\boldsymbol{\theta}}$ and $\boldsymbol{\theta}$ are in rad. The estimation errors of OGSBI, BCS and MUSIC methods are $-25.20$ dB, $-26.78$ dB and $-28.94$ dB, respectively. Since the mutual coupling effect has not been considered, the DOA estimation performance of these $3$ present methods almost the same. However, the DOA estimation error of the proposed SBLMC is $-49.31$ dB, which is significantly better than the present methods. 

Additionally, the DOA estimation performance of the proposed SBLMC method is compared with present methods in Fig.~\ref{compare}. The CS-based method (L1 norm) is proposed in~\cite{6657792,5393285}, where the DOA estimation problem is converted into a sparse reconstruction problem via $\ell_1$ minimization. The MUSIC-like method is proposed in~\cite{ce2017}, where the mutual coupling effect is considered in the MUSIC-like method. The Capon-like is proposed in~\cite{4655353} for the DOA estimation in MIMO radar systems. As shown in this figure, the proposed method achieves the best DOA estimation performance in the scenario with unknown mutual coupling effect by estimating all the unknown parameters iteratively in the SBL-based method. However, the disadvantage of the proposed method is higher computational complexity than these present methods.

We also show the DOA estimation performance with different signal-to-noise ratios (SNRs) in Fig.~\ref{SNR}. As shown in this figure, when $\text{SNR} \leq -10$ dB, all the methods cannot work well, and the performance is almost the same. When $\text{SNR} > -10$ dB, the DOA estimation performance of present methods including OGSBI, BCS and MUSIC cannot be improved, and the estimation errors are around $-28$ dB. However, with improving SNR, the estimation performance of SBLMC can also be improved, and the final estimation error can be lower than $-50$ dB with $\text{SNR} \geq 5$ dB. With the Cram\'{e}r-Rao  lower bound  (CRLB) in~\cite{6994289}, Fig.~\ref{SNR} also shows the corresponding CRLB of DOA estimation. As shown in this figure, when $\text{SNR} > 5$ dB, the proposed method can approach CRLB. In Fig.~\ref{SNR}, the curve ``SBLMC ($\ang{2}$)'' is the SBLMC method with the grid size being $\delta=\ang{2}$ and the curve ``SBLMC ($\ang{1}$)'' is that with the grid size being $\delta=\ang{1}$. From the curves, we can see that the estimation performance of SBLMC can approach the CRLB with the dense sampling grids ($\delta=\ang{1}$). Therefore, the reason why the estimation error cannot be further reduced (error floor) is that the spatial spectrum is discretized to find the peak values and the discretized grids cannot be infinitely small. The proposed SBLMC method can improve the estimation performance, but the improvement is limited by the grid size.

Then, we also show the mutual coupling effect on the DOA estimation in Fig.~\ref{MC}, where the mutual coupling effect between adjacent antennas is from $-15$ dB to $-2$ dB. With increasing the mutual coupling effect, the DOA estimation error of BCS is from $-34$ dB to $-28$ dB. Since the grid effect in the BCS method, decreasing the mutual coupling effect cannot further improve the estimation performance when the mutual coupling between adjacent antennas is less than $-8$ dB. However, for both OGSBI and MUSIC methods, decreasing the mutual coupling effect can decrease the estimation error from around $-25$ dB to around $-50$ dB. For the proposed methods, since the mutual coupling vectors $\boldsymbol{c}_T$ and $\boldsymbol{c}_R$ are estimated, the mutual coupling has limited effect on the DOA estimation performance, and the estimation error can be lower than $-50$ dB when the mutual coupling between adjacent antennas is less than $-2$ dB.

We show the DOA estimation performance with different grid sizes in Fig.~\ref{grid}, where the space between adjacent discretized angles $\delta$ is from $\ang{2}$ to $\ang{10}$. Since the BCS method has not considered the off-grid effect, the worst estimation performance is achieved in these $3$ methods. Both BCS and OGSBI methods have not considered the mutual coupling, so when the grid size $\delta$ is less than $\ang{6}$, the estimation performance cannot be improved. However, for the proposed SBLMC, with decreasing the gird size $\delta$ from $\ang{10}$ to $\ang{2}$, the estimation error can be decreased from $-8$ dB to $-50$ dB.

\begin{table}[!t]
	\renewcommand{\arraystretch}{1.3}
	\caption{Computational Time}
	\label{table3}
	\centering
	\begin{tabular}{cccc}
		\hline
		\textbf{Methods} & \textbf{Time (one iteration)} & \textbf{Number of iterations} & \textbf{Total time}\\
		\hline
		\textbf{SBLMC}&$4.12$ s &$139$ &$537.71$ s\\
		\textbf{OGSBI}&$2.66$ s &$146$ &$374.79$ s\\
		\textbf{BCS}& $0.17$ s &$147$ &$17.23$ s\\
		\textbf{MUSIC}& -- & -- &$80.31$ s\\
	\hline
	\end{tabular}
\end{table}

Finally, we compare the computational time of the proposed SBLMC method with that of the present $3$ methods in Table~\ref{table3}.
All the methods have not been further optimized to decrease the computational time. Without the additional simplifications, the computational complexities of SBLMC in Step~\ref{5}, Step~\ref{8} and Step~\ref{10}  are $\mathcal{O}(PU^2MN+UM^2N^2+U^3)$, $\mathcal{O}(PU^2MN+UM^2N^2+U^3)$ and $\mathcal{O}(U^3+PU^2MN)$, so the computational complexities of SBLMC can be obtained as $\mathcal{O}(U^3+PU^2MN+UM^2N^2)$ per iteration and an additional computational workload of order $\mathcal{O}(U^2M^3N^3)$ for initialization. To simplify the representation, with $U\geq MN$, the computational complexity of SBLMC can be approximated by $\mathcal{O}(PU^3)$. If $U<MN$ the computational complexity of SBLMC can be approximated by $\mathcal{O}(PUM^2N^2)$. Therefore, the proposed SBLMC algorithm has the same order of computational complexity with OGSBI~\cite{yang2013}. Since MUSIC algorithm is a continue domain method, we estimate the target DOAs by discretizing the detection range $[-\ang{80}, \ang{80}]$ into $1.6\times 10^6$ grids. As shown in this table, the BCS method is the fastest among all methods, since the detection angle is discretized by $\delta=\ang{2}$. The proposed SBLMC is comparable with the OGSBI method, but the DOA estimation performance is much better. The computational time of MUSIC method is determined by the length of discretized angles, and usually is a method with higher computational complexity than BCS. Therefore, the proposed SBLMC method can significantly improve the estimation performance in the MIMO radar system with both off-grid and mutual coupling effects with the acceptable computational complexity.

\section{Conclusions}\label{sec5}
We have investigated the DOA estimation problem in MIMO radar system with unknown mutual coupling effect in this paper. The off-grid problem in the CS-based sparse reconstruction method has also been considered concurrently to improve the DOA estimation performance. The novel sparse Bayesian learning with mutual coupling (SBLMC) method using EM has been proposed to estimate target DOAs.  Additionally, we have theoretically derived the prior distributions for all the unknown parameters including the variance vector of target scattering coefficients, the mutual coupling vectors, the off-grid vector and the noise variance. Simulation results confirm that the proposed SBLMC method outperforms the present DOA estimation methods in the MIMO radar system with the unknown mutual coupling effect. Additionally, the computational complexity of SBLMC is also acceptable. However, with the same characteristic of the sparse-based super-resolution methods, the minimum DOA separation of the proposed SBLMC method is limited by the radar aperture. Future work will focus on the optimization of MIMO radar system using the SBLMC method for DOA estimation.

\begin{figure*}
	\normalsize
	\vspace*{4pt}
	\begin{align}\label{ctLi}
	\frac{\partial \mathcal{L}(\boldsymbol{c}_\text{T})}{\partial \boldsymbol{c}_\text{T}}& = - \alpha_n P\left[\boldsymbol{c}^\text{H}\left(\sum_{p=0}^{U-1}\sum_{k=0}^{U-1}
	 \boldsymbol{\Upsilon}^\text{H}_p(\boldsymbol{\nu})\boldsymbol{\Upsilon}_k(\boldsymbol{\nu})
	 E_{x,k,p}
	 \right)
	 \begin{bmatrix}
	 	\boldsymbol{c}_\text{R}\otimes \boldsymbol{e}^M_0,\dots,\boldsymbol{c}_\text{R}\otimes \boldsymbol{e}^M_{M-1}
	 \end{bmatrix}\right]\notag \\
	 &+\sum^{P-1}_{p=0} \alpha_n [\boldsymbol{r}_p-\boldsymbol{\Upsilon}(\boldsymbol{\nu})(\boldsymbol{\mu}_p\otimes \boldsymbol{c})]^\text{H} \boldsymbol{\Upsilon}(\boldsymbol{\nu}) 
		\begin{bmatrix}
		\boldsymbol{\mu}_p\otimes \boldsymbol{c}_\text{R}\otimes \boldsymbol{e}^M_0 ,
			\dots,\boldsymbol{\mu}_p\otimes \boldsymbol{c}_\text{R}\otimes \boldsymbol{e}^M_{M-1} 
		\end{bmatrix}-\boldsymbol{c}^{H}_\text{T}\operatorname{diag}\{\boldsymbol{\vartheta_\text{T}}\}.
	\end{align}
	\begin{align}\label{nuLi}
\sum_{m=0}^{U-1}\nu_m\mathcal{R}\left\{\left(P\Sigma_{\text{X},u,m}+\sum^{P}_{p=0}
	 \mu^\text{H}_{p,m} \mu_{p,u}\right)\boldsymbol{c}^\text{H} \boldsymbol{\Xi}^\text{H}_m\boldsymbol{\Xi}_u\boldsymbol{c}
  \right\} =&\sum^{P}_{p=0}\mathcal{R}\left\{\left[\boldsymbol{r}_p
- \boldsymbol{\Psi}(\boldsymbol{\mu}_p\otimes \boldsymbol{c})\right]^\text{H}\boldsymbol{\Xi}_u \mu_{p,u}\boldsymbol{c}\right\}\notag\\
  &\quad -\sum_{m=0}^{U-1}\mathcal{R}\left\{P\Sigma_{\text{X},u,m}\boldsymbol{c}^\text{H} \boldsymbol{\Psi}^\text{H}_m\boldsymbol{\Xi}_u\boldsymbol{c}\right\}.
\end{align} 
		\hrulefill
\end{figure*}

\appendices

\section{The Derivation of Likelihood Function $\mathcal{L}(\boldsymbol{c}_\text{T}) $}\label{ct}
The likelihood function $\mathcal{L}(\boldsymbol{c}_\text{T}) $ can be rewritten as
\begin{align}	
\mathcal{L}&(\boldsymbol{c}_\text{T}) \propto 
	 -\alpha_n P \mathcal{G}_{1} (\boldsymbol{c}_\text{T})  -\sum^{P-1}_{p=0} \alpha_n \mathcal{G}_2(\boldsymbol{c}_\text{T}) - \mathcal{G}_3(\boldsymbol{c}_\text{T}),
	 \end{align}
where
\begin{align}
	\mathcal{G}_{1} (\boldsymbol{c}_\text{T})&\triangleq
	 \operatorname{Tr}\left\{
	 (\boldsymbol{I}_U\otimes \boldsymbol{c})^\text{H}
	\boldsymbol{\Upsilon}^\text{H}(\boldsymbol{\nu})
	 \boldsymbol{\Upsilon}(\boldsymbol{\nu})(\boldsymbol{I}_U\otimes \boldsymbol{c})\boldsymbol{\Sigma}_\text{X}\right\}, \\
	 \mathcal{G}_2(\boldsymbol{c}_\text{T}) &\triangleq \|\boldsymbol{r}_p-\boldsymbol{\Upsilon}(\boldsymbol{\nu})(\boldsymbol{\mu}_p\otimes \boldsymbol{c})\|^2_2,\\
	  \mathcal{G}_3(\boldsymbol{c}_\text{T})&\triangleq \sum^{M-1}_{m=0}  \vartheta_{\text{T},m} |c_{\text{T},m}|^2.
\end{align}

With the derivations of complex vector and matrix, $\frac{\partial \mathcal{G}_{1} (\boldsymbol{c}_\text{T}) }{\partial \boldsymbol{c}_\text{T}}$ is a row vector, and the $m$-th entry can be calculated as
\begin{align}
	&\left[\frac{\partial \mathcal{G}_{1} (\boldsymbol{c}_\text{T}) }{\partial \boldsymbol{c}_\text{T}}\right]_m = \operatorname{Tr}\left\{
\frac{\partial 
	 (\boldsymbol{I}_U\otimes \boldsymbol{c})^\text{H}
	\boldsymbol{\Upsilon}^\text{H}(\boldsymbol{\nu})
	 \boldsymbol{\Upsilon}(\boldsymbol{\nu})(\boldsymbol{I}_U\otimes \boldsymbol{c})\boldsymbol{\Sigma}_\text{X} }{\partial c_{\text{T},m}}\right\}. 
	 \end{align}
We can calculate 	
	 \begin{align}
	 & \frac{\partial 
	 (\boldsymbol{I}_U\otimes \boldsymbol{c})^\text{H}
	\boldsymbol{\Upsilon}^\text{H}(\boldsymbol{\nu})
	 \boldsymbol{\Upsilon}(\boldsymbol{\nu})(\boldsymbol{I}_U\otimes \boldsymbol{c})\boldsymbol{\Sigma}_\text{X} }{\partial c_{\text{T},m}}\notag\\
	 & = \frac{\partial 
	 (\boldsymbol{I}_U\otimes \boldsymbol{c})^\text{H}
	 }{\partial c_{\text{T},m}}\boldsymbol{\Upsilon}^\text{H}(\boldsymbol{\nu})
	 \boldsymbol{\Upsilon}(\boldsymbol{\nu})(\boldsymbol{I}_U\otimes \boldsymbol{c})\boldsymbol{\Sigma}_\text{X}\notag\\
	 &\qquad+(\boldsymbol{I}_U\otimes \boldsymbol{c})^\text{H}
	\boldsymbol{\Upsilon}^\text{H}(\boldsymbol{\nu})
	 \boldsymbol{\Upsilon}(\boldsymbol{\nu})\frac{\partial 
	 (\boldsymbol{I}_U\otimes \boldsymbol{c}) }{\partial c_{\text{T},m}}\boldsymbol{\Sigma}_\text{X}\notag\\ 
	 & = (\boldsymbol{I}_U\otimes \boldsymbol{c})^\text{H}
	\boldsymbol{\Upsilon}^\text{H}(\boldsymbol{\nu})
	 \boldsymbol{\Upsilon}(\boldsymbol{\nu})\left(\boldsymbol{I}_U\otimes   \frac{\partial
	 \boldsymbol{c}_\text{R}\otimes \boldsymbol{c}_\text{T} }{\partial c_{\text{T},m}}\right)\boldsymbol{\Sigma}_\text{X}\notag\\
	 & = (\boldsymbol{I}_U\otimes \boldsymbol{c})^\text{H}
	\boldsymbol{\Upsilon}^\text{H}(\boldsymbol{\nu})
	 \boldsymbol{\Upsilon}(\boldsymbol{\nu})\left(\boldsymbol{I}_U\otimes \boldsymbol{c}_\text{R}\otimes \boldsymbol{e}^{M}_m\right)\boldsymbol{\Sigma}_\text{X},
\end{align}	 
where $\boldsymbol{e}^M_m$ is a $M\times 1$ vector with the $m$-th entry being $1$ and  other entries being $0$. Therefore, the the $m$-th entry can be simplified as
\begin{align}
	\left[\frac{\partial \mathcal{G}_{1} (\boldsymbol{c}_\text{T}) }{\partial \boldsymbol{c}_\text{T}}\right]_m&=\boldsymbol{c}^\text{H}\left(\sum_{p=0}^{U-1}\sum_{k=0}^{U-1}
	 \boldsymbol{\Upsilon}^\text{H}_p(\boldsymbol{\nu})\boldsymbol{\Upsilon}_k(\boldsymbol{\nu})
	 \Sigma_{\text{X},k,p}
	 \right)\notag\\
	 &\qquad (\boldsymbol{c}_\text{R}\otimes \boldsymbol{e}^M_m),
\end{align}
and we finally have the derivation of $\mathcal{G}_{1} (\boldsymbol{\boldsymbol{c}_\text{T}})$ as
\begin{align}
	\frac{\partial \mathcal{G}_{1} (\boldsymbol{c}_\text{T}) }{\partial \boldsymbol{c}_\text{T}} &=\boldsymbol{c}^\text{H}\left(\sum_{p=0}^{U-1}\sum_{k=0}^{U-1}
	 \boldsymbol{\Upsilon}^\text{H}_p(\boldsymbol{\nu})\boldsymbol{\Upsilon}_k(\boldsymbol{\nu})
	 \Sigma_{\text{X},k,p}
	 \right)\\
	 &\qquad
	 \begin{bmatrix}
	 	\boldsymbol{c}_\text{R}\otimes \boldsymbol{e}^M_0,\boldsymbol{c}_\text{R}\otimes \boldsymbol{e}^M_1,\dots,\boldsymbol{c}_\text{R}\otimes \boldsymbol{e}^M_{M-1}
	 \end{bmatrix}.\notag
\end{align}

$\frac{\partial \mathcal{G}_{2} (\boldsymbol{c}_\text{T}) }{\partial \boldsymbol{c}_\text{T}}$  can be simplified as
\begin{align}
	\frac{\partial \mathcal{G}_{2} (\boldsymbol{c}_\text{T}) }{\partial \boldsymbol{c}_\text{T}}&=-[\boldsymbol{r}_p-\boldsymbol{\Upsilon}(\boldsymbol{\nu})(\boldsymbol{\mu}_p\otimes \boldsymbol{c})]^\text{H} \boldsymbol{\Upsilon}(\boldsymbol{\nu}) \frac{\partial \boldsymbol{\mu}_p\otimes \boldsymbol{c}}{\partial \boldsymbol{c}_\text{T}}\notag \\
		&=-[\boldsymbol{r}_p-\boldsymbol{\Upsilon}(\boldsymbol{\nu})(\boldsymbol{\mu}_p\otimes \boldsymbol{c})]^\text{H} \boldsymbol{\Upsilon}(\boldsymbol{\nu}) \\
		&\qquad \begin{bmatrix}
		\boldsymbol{\mu}_p\otimes \boldsymbol{c}_\text{R}\otimes \boldsymbol{e}^M_0 ,
			\dots,\boldsymbol{\mu}_p\otimes \boldsymbol{c}_\text{R}\otimes \boldsymbol{e}^M_{M-1} 
		\end{bmatrix}.\notag
\end{align}

$\frac{\partial \mathcal{G}_{2} (\boldsymbol{c}_\text{T}) }{\partial \boldsymbol{c}_\text{T}}$ can be simplified as
\begin{align}
	\frac{\partial \mathcal{G}_{2} (\boldsymbol{c}_\text{T}) }{\partial \boldsymbol{c}_\text{T}}=\boldsymbol{c}^{H}_\text{T}\operatorname{diag}\{\boldsymbol{\vartheta}_\text{T}\}.
\end{align}

Finally, with $\frac{\partial \mathcal{G}_{1} (\boldsymbol{c}_\text{T}) }{\partial \boldsymbol{c}_\text{T}}$, $\frac{\partial \mathcal{G}_{2} (\boldsymbol{c}_\text{T}) }{\partial \boldsymbol{c}_\text{T}}$ and $\frac{\partial \mathcal{G}_{3} (\boldsymbol{c}_\text{T}) }{\partial \boldsymbol{c}_\text{T}}$, the expression of $\frac{\partial \mathcal{L}(\boldsymbol{c}_\text{T})}{\partial \boldsymbol{c}_\text{T}}$ can be obtained in (\ref{ctLi}).

\section{The Derivation of Likelihood Function $\mathcal{L}(\boldsymbol{\nu}) $}\label{nuD}

The likelihood function $\mathcal{L}(\boldsymbol{c}_\text{T}) $ can be rewritten as
\begin{align}	
\mathcal{L}&(\boldsymbol{\nu}) \propto 
	 \sum^{P-1}_{p=0} \mathfrak{T}_1(\boldsymbol{\nu}) + \mathfrak{T}_2(\boldsymbol{\nu}),
	 \end{align}
where
\begin{align}
	\mathfrak{T}_{1} (\boldsymbol{\nu})&\triangleq \|\boldsymbol{r}_p-\boldsymbol{\Upsilon}(\boldsymbol{\nu})(\boldsymbol{\mu}_p\otimes \boldsymbol{c})\|^2_2, \\
	 \mathfrak{T}_2(\boldsymbol{\nu}) &\triangleq \operatorname{Tr}\{
(\boldsymbol{I}_U\otimes \boldsymbol{c})^\text{H}\boldsymbol{\Upsilon}^\text{H}(\boldsymbol{\nu})\boldsymbol{\Upsilon}(\boldsymbol{\nu})(\boldsymbol{I}_U\otimes \boldsymbol{c})\boldsymbol{\Sigma}_\text{X}\}.
\end{align}

$\frac{\partial \mathfrak{T}_1(\boldsymbol{\nu})}{\partial \boldsymbol{\nu}}$ can be obtained as
\begin{align}
&\frac{\partial \mathfrak{T}_1(\boldsymbol{\nu})}{\partial \boldsymbol{\nu}}  = -2\mathcal{R}\left\{[\boldsymbol{r}_p-\boldsymbol{\Upsilon}(\boldsymbol{\nu})(\boldsymbol{\mu}_p\otimes \boldsymbol{c})]^\text{H}\frac{\partial \boldsymbol{\Upsilon}(\boldsymbol{\nu})(\boldsymbol{\mu}_p\otimes \boldsymbol{c})}{\partial \boldsymbol{\nu}}\right\}\notag\\
& = -2\mathcal{R}\left\{[\boldsymbol{r}_p-\boldsymbol{\Upsilon}(\boldsymbol{\nu})(\boldsymbol{\mu}_p\otimes \boldsymbol{c})]^\text{H}  \boldsymbol{\Xi}(\operatorname{diag}\{\boldsymbol{\mu}_p\}\otimes \boldsymbol{c})
\right\}.
\end{align}

$\frac{\partial \mathfrak{T}_2(\boldsymbol{\nu})}{\partial \boldsymbol{\nu}} \in \mathbb{R}^{1\times U}$ is a row vector, and the $u$-th entry is
\begin{align}
&\left[\frac{\partial \mathfrak{T}_2(\boldsymbol{\nu})}{\partial \boldsymbol{\nu}}\right]_u  = \operatorname{Tr}\left\{\frac{\partial 
(\boldsymbol{I}_U\otimes \boldsymbol{c})^\text{H}\boldsymbol{\Upsilon}^\text{H}(\boldsymbol{\nu})\boldsymbol{\Upsilon}(\boldsymbol{\nu})(\boldsymbol{I}_U\otimes \boldsymbol{c})\boldsymbol{\Sigma}_\text{X}}{\partial \nu_u}\right\}\notag\\
&\quad = \operatorname{Tr}\left\{
	\begin{bmatrix}
		\boldsymbol{0},(\boldsymbol{I}_U\otimes\boldsymbol{c}^\text{H})\boldsymbol{\Upsilon}^\text{H}(\boldsymbol{\nu})\boldsymbol{\Xi}_u\boldsymbol{c},\boldsymbol{0}
	\end{bmatrix}\right\}\notag\\
	&\qquad 
	+\operatorname{Tr}\left\{\begin{bmatrix}
		\boldsymbol{0},(\boldsymbol{I}_U\otimes\boldsymbol{c}^\text{H})\boldsymbol{\Upsilon}^\text{H}(\boldsymbol{\nu})\boldsymbol{\Xi}_u\boldsymbol{c},\boldsymbol{0}
	\end{bmatrix}^\text{H} \boldsymbol{\Sigma}_\text{X}
\right\}\notag\\
&\quad=2\mathcal{R}\left\{\sum_{m=0}^{U-1}\boldsymbol{c}^\text{H}\boldsymbol{\Upsilon}^\text{H}_{m}(\boldsymbol{\nu})\boldsymbol{\Xi}_u\boldsymbol{c}\Sigma_{\text{X},u,m}\right\}.
\end{align} 
Therefore, $\frac{\partial \mathfrak{T}_2(\boldsymbol{\nu})}{\partial \boldsymbol{\nu}} $ can be simplified as
\begin{align}
\frac{\partial \mathfrak{T}_2(\boldsymbol{\nu})}{\partial \boldsymbol{\nu}}  = 2\mathcal{R}\left\{\operatorname{diag}\left\{
\boldsymbol{\Sigma}_\text{X}(\boldsymbol{I}_U\otimes\boldsymbol{c})^\text{H}\boldsymbol{\Upsilon}^\text{H}(\boldsymbol{\nu})\boldsymbol{\Xi}(\boldsymbol{I}_U\otimes \boldsymbol{c})\right\}^\text{T}
	\right\}.
\end{align}  

Therefore, with $\frac{\partial \mathcal{L}(\boldsymbol{\nu})}{\partial \nu_u}=0$, we can obtain the equation (\ref{nuLi}) to obtain $\boldsymbol{\nu}$.


\bibliographystyle{IEEEtran}
\bibliography{IEEEabrv,References}

\begin{IEEEbiography}[{\includegraphics[width=1in,height=1.25in,clip,keepaspectratio]{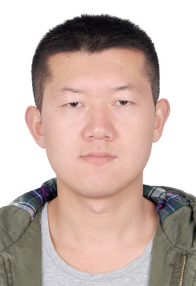}}]{Peng Chen (S'15-M'17)}
was born in Jiangsu, China in 1989. He received the B.E. degree in 2011 and the Ph.D. degree in 2017, both from the School of Information Science and Engineering, Southeast University, China. From Mar. 2015 to Apr. 2016, he was a Visiting Scholar in the Electrical Engineering Department, Columbia University, New York, NY, USA.

He is now an associate professor at the State Key Laboratory of Millimeter Waves, Southeast University. His research interests include radar signal processing and millimeter wave communication.

\end{IEEEbiography}

\begin{IEEEbiography}[{\includegraphics[width=1in,height=1.25in,clip,keepaspectratio]{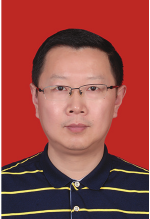}}]{Zhenxin Cao (M'18)}
   was born in May 1976. He received the M. S. degree  in 2002 from Nanjing University of Aeronautics and Astronautics, China, and the Ph.D. degree in 2005 from  the School of Information Science and Engineering, Southeast University, China. From 2012 to 2013, he was a Visiting Scholar in North Carolina State University. 
	
	Since 2005, he has been with the State Key Laboratory of Millimeter Waves, Southeast University, where he is a Professor. His research interests include antenna theory and application. 

\end{IEEEbiography}

\begin{IEEEbiography}[{\includegraphics[width=1in,height=1.25in,clip,keepaspectratio]{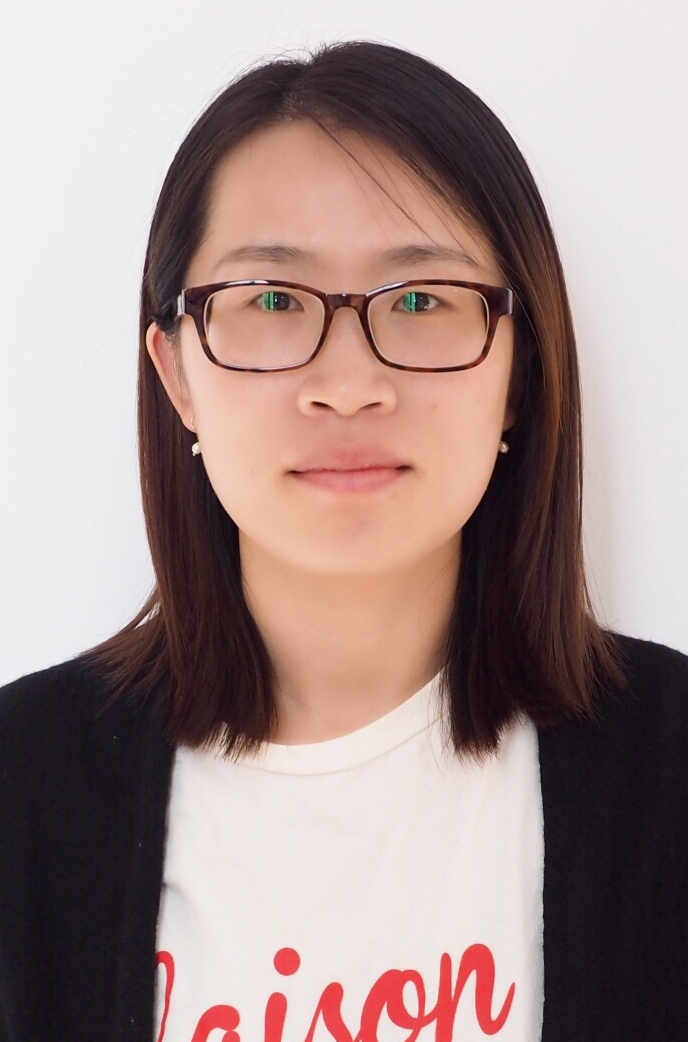}}]{Zhimin Chen (M'17)} was born in Shandong, China, in 1985. She received the Ph.D. degree in information and communication engineering from the School of Information Science and Engineering, Southeast University, Nanjing, China in 2015. Since 2015, she has been with Shanghai Dianji University, Shanghai, China. Her research interests include array signal pro-cessing and Millimeter-Wave communications.
	
\end{IEEEbiography}

\begin{IEEEbiography}[{\includegraphics[width=1in,height=1.25in,clip,keepaspectratio]{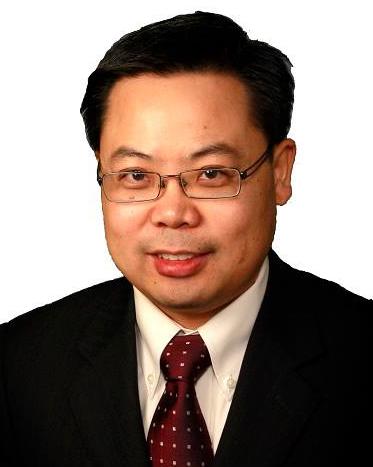}}]{Xianbin Wang (S'98-M'99-SM'06-F'17)} is a Professor and Tier-I Canada Research Chair at Western University, Canada. He received his Ph.D. degree in electrical and computer engineering from National University of Singapore in 2001.

Prior to joining Western, he was with Communications Research Centre Canada (CRC) as a Research Scientist/Senior Research Scientist between July 2002 and Dec. 2007. From Jan. 2001 to July 2002, he was a system designer at STMicroelectronics, where he was responsible for the system design of DSL and Gigabit Ethernet chipsets.  His current research interests include 5G technologies, Internet-of-Things, communications security, machine learning and locationing technologies. Dr. Wang has over 300 peer-reviewed journal and conference papers, in addition to 26 granted and pending patents and several standard contributions.

Dr. Wang is a Fellow of Canadian Academy of Engineering, a Fellow of IEEE and an IEEE Distinguished Lecturer. He has received many awards and recognitions, including Canada Research Chair, CRC President’s Excellence Award, Canadian Federal Government Public Service Award, Ontario Early Researcher Award and five IEEE Best Paper Awards. He currently serves as an Editor/Associate Editor for IEEE Transactions on Communications, IEEE Transactions on Broadcasting, and IEEE Transactions on Vehicular Technology and He was also an Associate Editor for IEEE Transactions on Wireless Communications between 2007 and 2011, and IEEE Wireless Communications Letters between 2011 and 2016. Dr. Wang was involved in many IEEE conferences including GLOBECOM, ICC, VTC, PIMRC, WCNC and CWIT, in different roles such as symposium chair, tutorial instructor, track chair, session chair and TPC co-chair.

\end{IEEEbiography}

\end{document}